\documentclass[12pt]{article}

\usepackage{amsmath,amsfonts}
\usepackage{graphicx}
\usepackage{amsmath}
\usepackage{hyperref}
\usepackage{amsthm}
\usepackage{bbm}
\usepackage{amssymb}
\usepackage{hyperref}
\hypersetup{
    colorlinks=true,
    linkcolor=blue,
    filecolor=magenta,
    urlcolor=cyan,
}
\usepackage{eucal}

\newtheorem{theorem}{Theorem}
\newtheorem*{theorem*}{Theorem}

\newtheorem{claim}{Claim}

\newtheorem{lemma}{Lemma}

\newtheorem{proposition}{Proposition}
\newtheorem{assumption}{Assumption}
\newtheorem{defn}{Definition}
\newtheorem{example}{Example}
\newtheorem{corollary}{Corollary}

\theoremstyle{remark}
\newtheorem{remark}{Remark}

\DeclareMathOperator*{\argmin}{arg\,min}

\def\b{\beta}
\def\d{\delta}
\def\dk{\Delta(K)}

\def\o{\omega}
\def\k{\kappa}

\def\g{\gamma}
\def\s{\sigma}

\def\S{\Sigma}
\def\t{\tau}
\def\l{\ell}
\def\k{\kappa}
\def\p{\pi_M}

\def\N{\mathbb{N}}
\def\Y{\mathcal{Y}}
\def\U{\mathcal{U}}
\def\V{\mathcal{V}}
\def\W{\mathcal{W}}

\def\G{\Gamma}
\def\R{\mathcal{R}}
\def\D{\Delta}
\def\x{\xi}
\def\1{\mathbbm{1}}

\newcommand{\headings}[1]{\vskip .3cm \noindent \textbf {#1} \hskip .4cm }
\newcommand{\ignore}[1]{}

\begin{document}
\title{Markovian Persuasion with Stochastic Revelations}


\author{Ehud Lehrer\footnote{Department of Economics, Durham University, Mill Hill Lane, Durham DH1 3LB, UK. Email:
ehud.m.lehrer@durham.ac.uk; Lehrer acknowledges the support of  ISF 591/21 and DFG KA 5609/1-1. }\ \ and Dimitry Shaiderman\footnote{Department of Mathematics, Hebrew University of Jerusalem, Israel, email: dima.shaiderman@gmail.com. The authors are greatly indebted to the three anonymous referees of \emph{GEB} for their insightful and detailed comments, suggestions, and overall assistance, which significantly improved and extended the scope and exposition of the current work. In addition, the authors would like to deeply thank Andrei Iacob and Eilon Solan for their care, help, and contributions to this work.} }

\maketitle
\thispagestyle{empty}
\begin{abstract}

\footnotesize{This paper examines information transmission games where the sender
knows the realizations of states from a Markov process, but his informational
advantage is counteracted by outside, random revelations to the receiver. For very patient players, we characterize the sender’s value of the game with outside revelations in terms of the value without outside revelations. Through our characterization, we identify that, in contrast with the sender-receiver
games with fixed states, the sender may benefit from being patient: his
discounted value of the game may increase as the discount factor grows.}
\vskip .3cm

\noindent{\footnotesize \textbf{Keywords: } Markovian persuasion, Markov chains, Stochastic revelations, Monotonic trajectories, Dynamic information provision. }

\vskip .3cm

\noindent{\footnotesize \textbf{JEL Codes:} D72, D82, D83, K40, M31.}
\end{abstract}

\newpage
\setcounter{page}{1}

\section{Introduction}

Information control is a valuable asset to individuals interacting over time. Whoever controls it has the power to persuade those who lack it to take particular actions, through strategic disclosures. A lot of attention has been paid to the dynamic persuasion problem lately, but most of the literature examines games in which an uninformed player, the receiver, can only obtain information from an informed player, the sender. This paper investigates how the sender’s value from persuasion is affected by the existence of exogenous learning of the receiver.


\textit{Model.} The model at hand considers a receiver, who obtains information about the current state of a discrete-time Markov chain with Markov matrix $M$ from two separate sources. The first is the sender, who observes the chain in real-time, and at every stage must transmit a signal to the receiver from a pre-committed information provision policy. The second source of information is described by \textit{stochastic revelations}; at each stage, after obtaining information from the sender, the receiver learns about the occurrence or lack thereof of a revelation. Such a revelation, independent of the evolution of the Markov chain and the sender's signals, occurs at every stage with fixed probability $x \in [0,1]$, titled the \textit{revelation rate}. Upon a revelation, the current state of the Markov chain is disclosed to the receiver. 


Both sources of information allow the receiver to sequentially update her beliefs about the current state of the Markov chain. At each stage, the order in which she obtains her information is of importance, as her \textit{action} at that period should come after obtaining the current signal of the sender, and before learning of the occurrence of a revelation. The receiver's action policy is a function of her belief and is assumed to be known to the sender. The stage action of the receiver, together with the current state of the Markov chain, determines the stage payoff of the sender. The total payoff of the sender is his expected $\d$-discounted sum of the stage payoffs, where $\d$ denotes the discount factor. 


The occurrence or lack thereof of stochastic revelations is also assumed to be observed by the sender. In practice, this assumption means that although the sender has no monopoly on the belief process of the receiver, he still can track the latter. This is crucial as it enables the sender to know which action will be assigned to each signal he transmits. In that view, the sender's optimization problem is to optimally influence a belief process, which he has no full control over.   


\textit{Economic Applicability.} The model at hand may be applied to a case where the sender is used to model a large company or government agency, that tries to persuade the public to take a certain action. In such cases, each individual affiliated with the former might leak valuable information to the public. Thus, the sender as an entity may consider a scenario where at each time period, the true state of its affairs is stochastically revealed to the public by one of its affiliates.   


\textit{Main Results.} The first main result of the paper, stated in Theorem \ref{Thm1}, concerns the optimal behavior of a very patient sender (i.e., the discount factor $\d$ approaches $1$). Its main findings are twofold. First, for every positive revelation rate $x>0$, the limit of the $\d$-discounted values as $\d$ tends to $1$ may be expressed in terms of an average of the $(1-x)$-discounted values of the game with \textit{no revelations}  computed at a special family of priors.\footnote{By games with no revelations, we refer to those in which the revelation rate equals $0$.} This family consists of the rows of the Markov matrix $M$, describing the possible beliefs of the receiver at the start of stages that follow revelations. Second, there exists a strategy for the sender, which approximates this limit, for any discount factor $\d$ large enough.    


The main practical implication of Theorem \ref{Thm1}, detailed in Subsection \ref{Subsec Rev Effect}, addresses the effect of the revelations on the sender. It turns out that, in some instances, no matter how small the revelation rate $x$ is, and how large the patience level is (large values of $\d$), the existence of revelations reduces the sender's $\d$-discounted values compared to the scenario where revelations do not occur.

The second main result of the paper concerns the effect of the discount factor $\d$ on the $\d$-discounted values of the game with no revelations. Theorem \ref{Thm Mono 1} finds that whenever $M$ is irreducible, a certain average of the $\d$-discounted values of such games, computed at the rows of $M$, must increase as $\d$ grows. Theorem \ref{Thm Mono 2} generalizes Theorem \ref{Thm Mono 1} to infinitely many sets of priors, whose convex hull contains that of the rows of $M$, showing that certain averages of the corresponding $\d$-discounted values must also increase with $\d$. Theorems \ref{Thm Mono 1} and \ref{Thm Mono 2}, which address the game with no revelations, were made possible by studying games with revelations.

The main implication of Theorems \ref{Thm Mono 1} and \ref{Thm Mono 2} is that in the absence of revelations, there are infinitely many priors from which the sender would benefit by being patient. Namely, for any two discount factors, there exist infinitely many priors starting from which the discounted value with the higher discount exceeds that with the lower discount. (see Corollary \ref{Corol Mono 1 and 2}). Contrary to the well-documented phenomena of the \textit{decrease} of discounted values as the discount grows for certain models of repeated games, the current implication sheds new light on the effect of patience on dynamic repeated games. 


\textit{Key Heuristics behind Theorem \ref{Thm1}.} The proof of Theorem \ref{Thm1} capitalizes on an embedding of the model in a general class of Markov decision processes (MDPs) (see Section \ref{Section General MDP}). Such embedding allows to switch attention from $\d$-discounted payoffs to $N$-Ces\'aro average ones. In the latter, the main focus is on the \textit{sum of payoffs} in stages occurring between any two successive revelations. 


Along such stages, the sender participates in a \textit{random-duration game with no revelations}, whose duration follows a geometric distribution with parameter $x$. The play in such a game, admitting a prior within the rows of $M$, is independent of the play in all previous and future stages, as after each revelation, the information of the receiver is rebooted. This reboot is due to the unique nature of the revelations; the former being independent of all other events in the game.


As it turns out, the expected sum of stage payoffs in such random-duration games with no revelations, admits a formula in terms of the $(1-x)$-discounted payoff for games with no revelations (see Lemma \ref{Random Duration Value Lemma} in Subsection \ref{Subsection Random Duration Valuation}). Consequently, the sender's optimal behavior under the $N$-Ces\'aro average payoff is to play, based on the prior row of $M$, his optimal strategy in the $(1-x)$-discounted game with no revelations, in between any two successive revelations. The computation of the limit payoff (as $N$ tends to infinity) under this strategy requires a stochastic analysis intended to assess the limit frequency with which each row of $M$ serves as a prior after a revelation (given in Section \ref{Proof of Uniform and Asym Values}). Such analysis employs fundamental tools from probability theory, including the Ergodic Theorem for irreducible Markov chains.  A detailed survey of the heuristics behind the proof of Theorem \ref{Thm1} is given in Subsection \ref{subsec:heuristics 1}.



\textit{Key Heuristics behind Theorems \ref{Thm Mono 1} and \ref{Thm Mono 2}}. The proof of Theorem \ref{Thm Mono 1} builds on the following informational coupling argument. Whenever a revelation did not occur (with probability $1-x$), knowing this, the sender can ``disclose" to the receiver the state of the Markov chain with probability $(y-x)/(1-x)$, where $y \in (x,1]$. Such coupling would produce at any stage a revelation at rate $y>x$. Thus, the game with revelation rate $y>x$, can be embedded in that with revelation rate $x$. In particular, each $N$-Ces\'aro value will decrease as a function of the revelation rate. Thus, the limit of such decreasing functions (as $N$ tends to infinity), described in Theorem \ref{Thm1}, will also decrease. Making the change of variable $\d = 1-x$, will flip the monotonicity to establish the existence of an average of $\d$-discounted values that increase with the discount factor. 



The main technical challenge is to formalize the ``disclosing" part in terms of the signals available to the sender. Such formalization, given in Section \ref{Sec Proof of Mono 1}, requires studying recursive formulas for the $N$-Ces\'aro values, as well as employing fundamental tools from Convex Theory such as  Carath\'{e}odory's Theorem.


The generalization of Theorem \ref{Thm Mono 1}, stated in Theorem \ref{Thm Mono 2}, requires a generalization of the model setup. In essence, instead of fully revealing, the newly considered stochastic revelations give partial information about the realized state of the Markov chain; they do so by splitting the receiver's beliefs into a pre-specified set of beliefs whose convex hull contains that of the rows of $M$. After the formal introduction of such a model (see Subsection \ref{SubSec Proof Mono 2}), the proof of the respective monotonicity result follows from suitable modifications of the proofs of Theorems \ref{Thm Mono 1} and \ref{Thm Mono 2}. A detailed survey of the heuristics behind the proof of Theorems \ref{Thm Mono 1} and \ref{Thm Mono 2} is given in Subsections \ref{Subsec Mono Heuristics 1} and \ref{Subsec Mono Heuristics 2}, respectively.



\textit{Organization of the paper.} The formal description of the model under consideration is given in Section \ref{sec: model}. Section \ref{sec: main results} is devoted to the exposition and statement of the main results. Thereafter, a detailed survey of the mathematical heuristics behind the main results appears in Section \ref{sec:heuristics}. Section \ref{sec: application to repeated games} applies the tools and techniques considered in the paper to a model of zero-sum repeated games with changing states. In Section \ref{Section General MDP} we embed the different models under consideration in a general MDP framework, which allows for a robust analysis across all models. The formal proofs are given in Sections \ref{Proof of Uniform and Asym Values}, \ref{Sec Proof of Mono 1}, and \ref{sec: proof mono 2}. 


\subsection{Related literature}

The closest work to the present one is that by Lehrer and Shaiderman \cite{MP}. The former studies a particular case of the current model, in which there are no revelations, or equivalently, where the revelation rate is $0$. The research in \cite{MP} focuses on the convergence of the $\d$-discounted values as $\d$ tends to $1$ as well. More specifically, it characterizes the cases in which such values converge to their upper bound, in the case where $M$ is irreducible and aperiodic (see Theorem 2 in \cite{MP}).\footnote{This upper bound is the solution to the static persuasion problem of Gentzkow and Kamenica \cite{Kamenica} at the unique stationary distribution of $M$.} As it turns out, when such upper bound is achieved, it also majorizes for any positive revelation rate $x>0$ the limit of values described in Theorem \ref{Thm Mono 1}. The latter finding allows us to understand better how the sender's payoff is affected by stochastic revelations (see Subsection \ref{Subsec Rev Effect}).


The revelation-less model in Lehrer and Shaiderman \cite{MP} sets a general framework, inspired by several previously studied instances of optimal information provision problems in a Markovian environment. Such problems include those studied in Renault, Solan, Vieille \cite{Solan}, Ely \cite{Ely}, Farhadi and Teneketzis \cite{Farhadi}, and Ashkenazi-Golan et al. \cite{Galit}. Such works showcased that even when the sender fully controls the flow of the receiver's information, he does not always benefit from the greedy strategy. However, when adding stochastic revelations, the current research establishes that the sender should always be greedy in the following sense: under the expected $N$-Ces\'aro average payoff, he should always maximize his expected payoff between any two successive revelations. 


Additional works dealing with a receiver whose information is not fully controlled by the sender include Bizzotto, R\"{u}diger, and Vigier \cite{Bizzotto} and Lorecchio \cite{Lorecchio}. In the former, a sender tries to sequentially persuade a receiver to take a particular action, and commits to an information disclosure policy, but there is a possibility of the receiver exogenously obtaining information about the state in later stages of the game. In the latter, the dynamic persuasion problem consists of a sequence of myopic receivers, each obtaining private information and, in addition, observing the actions of the past receivers (but not their private information). One difference between these two works and the previously mentioned ones is that the setup is static, that is, the underlying state does not
evolve over time.


In summary, the current work belongs to the literature on dynamic Bayesian persuasion, where the state evolves stochastically over time. Its novelty is in the showcase of new ways in which the tools of dynamic Bayesian persuasion can be employed to characterize communication in a class of games in which the sender is not an informational monopolist. The specific model, techniques, and results can be viewed as a step toward the analysis of more general dynamic persuasion problems in which the sender possesses only partial control over the information available to the receiver.


\subsection{Notations} 

For every countable set $A$, $\Delta(A)$ denotes the set of probability measures over $A$. For every $p \in \Delta(A)$ and $a \in A$, $p^a$ denotes the probability mass assigned by $p$ to $a$. For any subset $C \subseteq A$, $p(C)$ denotes the probability assigned to the event $C$ by $p$. Moreover, for every $a \in A$, let $\d_a \in \Delta(A)$ denote the Dirac measure supported on $a$.

If $A$ is finite, and $M$ is a matrix of dimensions $\vert A \vert \times \vert A \vert$, then for every $p \in \Delta(A)$,  $pM$ denotes the $M$\textit{-shift} of $p$; that is, $pM$ is the matrix multiplication of $p$ with $M$, where $p$ is viewed as a row vector of length $\vert A \vert$. In the case where $M$ is a stochastic matrix, $(pM)^a$ denotes the probability mass assigned to $a \in A$ by the probability measure $pM \in \Delta(A)$.

For any set of vectors $w_1,...,w_n$ in $\mathbb{R}^d$, $conv \{w_1,...,w_n\}$ denotes the convex hull spanned by those vectors. For any function $f:D \to \mathbb{R}$, where $D \subseteq \mathbb{R}^d$ is a compact convex domain, $(\text{Cav}\, f)$ denotes the \textit{concavification} of $f$, i.e., 
\begin{align*}
(\text{Cav}\, f)(x) := \inf \{h(x)\,:\, h:D \to \mathbb{R} \text{\, is concave}, f \leq h \}, \quad \forall x \in D.
\end{align*}
That is, $(\text{Cav}\, f)$ is the smallest concave function that majorizes $f$.

\section{The model}\label{sec: model}

We consider a two-player repeated game between a \emph{sender} and a \emph{receiver}, called \emph{Markovian persuasion game with stochastic revelations}. The roles of the players remain fixed throughout the different stages of the game.

In order to formally define the game, we need the following notations and assertions.
\begin{itemize}
  \item $K=\{1,...,k\}$ is a finite set of states.
  \item $S$ is a finite set of signals available to the sender, whose cardinality is at least $k$.
  \item $B$ is a finite set of actions available to the receiver.
  \item  $(X_n)_{n \geq 1}$ is a Markov chain over $K$ with prior probability $p \in \Delta(K)$ and a transition rule given by the stochastic matrix $M$. 
  \item $(Z_n)_{n \geq 1}$ is a sequence of i.i.d.\ Bernoulli trials. The success probability of each $Z_n$ is denoted by $x \in [0,1]$. The parameter $x$ is called the \emph{revelation rate}. 
  \item We assume that $(X_n)_{n \geq 1}$ and $(Z_n)_{n \geq 1}$ are defined on the probability space $(\Omega, \mathcal{F},P)$. In addition, we assume that for every $n\geq 1$, $Z_n$ is independent of $(X_n)_{n \geq 1}$.
\end{itemize}

The above list is commonly known among the players in our repeated game, which we denote by  $MP(x)$. The game $MP(x)$ is played in stages, and the timeline of each stage $n\geq 1$ consists of the following sequence of events:
\begin{itemize}
    \item The sender observes the realized state $x_n \in K$ of $X_n$.
    \item A signal $s_n \in S$ is transmitted by the sender to the receiver.   
    \item The receiver takes an action $b_n \in B$, which together with $x_n$ determines the payoffs of the sender and the receiver, given by the utilities $v(x_n,b_n)$  and $w(x_n,b_n)$, respectively. 
    \item The sender is informed of the receiver's action.
    \item Both players observe the realized value $z_n$ of the Bernoulli trial $Z_n$. The state $x_n$ is made public if and only if the trial is successful ($z_n=1$). In this case, we say that a \emph{revelation} occurred. 
    \item The game proceeds to stage $n+1$.
\end{itemize}

\begin{remark}
    In Remark \ref{Remark 2} appearing in Subsection \ref{SubSec Proof Mono 2} we address the case where at each stage $n\geq 1$, the revelation, or lack thereof,  occurs right before the receiver takes his action $b_n$. In terms of the above timeline, this captures the case where the fifth item in the above list moves up, and occurs right after the second item. 
\end{remark}

\subsection{Signaling policies}\label{subsec Model Probability space}

A \emph{signaling policy} $\sigma$ of the sender in $MP(x)$ is described by a sequence of stage strategies $(\s_n)_{n \geq 1}$, where $\s_n : (K \times S \times \{0,1\})^{n-1}\times K \to \Delta(S)$. The transmitted signal $s_n$ is drawn from the lottery $\s_n$, which may depend on the following information of the sender: the past states $x_1,...,x_{n-1}$, past signals $s_1,...,s_{n-1}$, and past Bernoulli trial outcomes $z_1,...,z_{n-1}$, together with the current state $x_n$. The reasoning for assuming that the signal $s_n$ is independent of the receiver's past actions $b_1,...,b_{n-1}$ will be made clear in Subsection \ref{SubSec Rec-Policy}. Denote $\S$ to be the space of all signaling policies.

The \emph{commitment assumption} requires the sender to make his signaling policy $\sigma \in \Sigma$ known to the receiver prior to the first stage of $MP(x)$. By doing so, the sender allows the receiver to assign probabilities to various events along the game. Formally, by Kolmogorov's Extension Theorem, each signaling policy $\s \in \S$ together with $(X_n)_{n \geq 1}$ and $(Z_n)_{n \geq 1}$ induces a unique probability measure $P^p_{x,\s}$ on the space $\Y = (K \times S \times \{0,1\})^{\N}$. This measure is determined by the following laws:
\begin{multline*}
P^p_{x,\s} (x_1,s_1,z_1,...,x_n,s_n,z_n) = \left(p(x_1)\prod_{i=1}^{n-1}M_{x_i,x_{i+1}}\right)\times \\ \left(\prod_{i=1}^n \s_i(x_1,s_1,z_1,...,x_{i-1},s_{i-1},z_{i-1},x_i)(s_i)\right) \times \left( \prod_{i=1}^n x^{z_i} (1-x)^{1-z_i}\right).
\end{multline*}

\subsection{Posterior beliefs}\label{subsection model posteriors}

For each stage $n\geq 1$, we introduce the random signal $R_n$ with values in $K \cup \{\emptyset \}$. It is defined by $R_n=X_n$ on $\{Z_n=1\}$ and $R_n=\emptyset$ on $\{Z_n=0\}$. It describes the receiver's additional information upon observing the $n$'th trial $Z_n$. As before, the realized value of $R_n$ is denoted by $r_n$.

The $n$'th posterior of the receiver, denoted $p_n$, is a random variable taking values in $\Delta(K)$, defined by
\begin{equation}\label{posterior sequence definition}
p_n^{\l} = P^p_{x,\s}\left(X_n=\l\,|\, s_1,r_1,...,s_{n-1},r_{n-1},s_n\right), \quad \forall \l \in K.
\end{equation}
That is, $p_n$ describes the posterior belief of the receiver on the current state of $X_n$, precisely at the moment at which she is required to take her action $b_n \in B$. This belief is affected by all information available to her leading to this moment, namely, the policy $\s$, the past signals, and the Bernoulli trial outcomes she observed, together with today's signal $s_n$.  

The \textit{beliefs process} of the receiver is given by the stochastic process $(p_n)_{n\geq 1}$.

\subsection{The receiver's policy}\label{SubSec Rec-Policy}

An important modeling assumption asserts that the sender determines the actions of the receiver using his signaling policy. Namely, at each stage $n\geq 1$, the receiver's action for that stage is determined by the value of his current posterior $p_n$, which is itself determined by the sender’s signaling policy. In formal terms, consider a mapping $\theta: \Delta(K) \to B$, satisfying: (i) $\theta(q) \in \text{arg\, max}_B \sum_{\l \in K} q^{\l} w(\l,b)$ for every $q \in \Delta(K)$ and (ii) $\sum_{\l \in K} q^{\l} v(\l,\theta(q)) \geq \sum_{\l \in K} q^{\l} v(\l,b)$ for every $q \in \Delta(K)$ and every $b \in \text{arg\, max}_B \sum_{\l \in K} q^{\l} w(\l,b)$. Then, at any stage $n\geq 1$ of $MP(x)$, the receiver's $n$'th action $b_n$ is taken to equal $\theta (p_n)$. We remark that Condition (i) implies that the receiver is myopic, whereas assertion (ii) requires the receiver to break ties in favor of the sender.


From a modeling standpoint, the receiver's strategic behavior discussed above may fit several instances. First, if we think of the receiver as a sequence of transitory short-lived receivers (e.g., Jackson and Kalai \cite{Jackson Kalai}) playing sequentially in a recurring game and having social memory of the information provided (as is often the case in online markets), then our assumption corresponds to the case where each receiver acts myopically in his turn. Another instance is where our single receiver embodies a scenario where, at each stage, a particular receiver is chosen from a large population of anonymous receivers. Such an instance fits a model of a political party or a media outlet (the sender) trying to persuade the general public.

\subsection{The sender's optimization problem}

The sender aims to maximize his expected $\delta$-discounted stream of payoffs across all possible signaling policies. A formal expression for the latter may be written as follows. First, define $u:\Delta(K) \to \mathbb{R}$ by
\begin{align}\label{u-definition}
    u(q) := \sum_{\l \in K} q^{\l} \, v(\l,\theta (q)), \quad \forall q \in \Delta(K).
\end{align}
Note that by the tie-breaking rule, $u$ is upper semi-continuous. Then, we associate with each prior $p \in \Delta(K)$ and signaling policy $\s \in \Sigma$, the $\delta$-discounted payoff, denoted $\gamma^x_{\delta} (p, \s)$, which is given by
\begin{equation*}
    \gamma_{\delta}^x (p, \s) = E^p_{x,\s}\left(\sum_{n=1}^{\infty} (1-\d)\d^{n-1} u(p_n)\right).
\end{equation*}
The sender's optimization problem is to calculate the value $v^x_{\delta} (p) = \max_{\sigma \in \Sigma} \gamma_{\delta}^x (p, \s)$, and describe a strategy $\s \in \S$ which achieves this maximum\footnote{The maximum is attained as the target function is upper semicontinuous and the domain of maximization is compact}.

\subsection{Relation to the Markovian persuasion model without revelations}

The Markovian persuasion model studied in Lehrer and Shaiderman \cite{MP} can be viewed as a particular case of the above model. In the former, there are no revelations, and the information of the receiver is solely generated by the sender's signaling policy. With the exception of this difference, the description of both models coincides. 

As we deal with a maximization problem of expected payoffs, from a value standpoint, the Markovian persuasion game, denoted $\Gamma (p)$, is identified with $\Gamma_0 (p)$, in which revelation occurs with zero probability. Thus, $v_{\d} (p) = v^0_{\d}(p)$, for every $\d$, where $v_{\d} (p)$ denotes the $\delta$-discounted value of $\Gamma(p)$. 

The similarities and differences between the two models give rise to several natural questions. One such question is how sensitive the sender is to the leakage of information occurring due to the stochastic revelations. In mathematical terms, this question may be addressed by comparing and quantifying the differences between $v_{\d}(p)$ and $v^x_{\d}(p)$ when $x>0$. Another natural question refers to the effect of the patience level, described by $\delta \in [0,1)$, on the sender. In particular, can one quantify the limiting behavior of  $v^x_{\d}(p)$, as $\d \to 1^{-}$, i.e., when the sender becomes arbitrarily patient? Should this limiting behavior be comparable with the limiting behavior of $v_{\d}(p)$? The analysis in the paper 
 strives to provide detailed answers to the mentioned questions.

\section{The main results}\label{sec: main results}
\subsection{The existence of asymptotic and uniform values}

Our first main result studies the limiting behavior of $v^x_{\d}(p)$, as $\d \to 1^{-}$, with respect to two classical solution concepts, known as the \textit{asymptotic} and \textit{uniform} values:

\begin{defn}\label{Definition Uniform Value}
Let $x \in [0,1]$.
\begin{itemize} 
    \item[\emph{1.}] The game $MP(x)$ admits an \emph{asymptotic value at $p$} if $\lim_{\d \to 1^-} v^x_{\d}(p)$ exists for every $p \in \dk$. In such a case, we denote the asymptotic value by $v_{\infty}^x (p)$.
        \item[\emph{2.}] The game $MP(x)$ admits a \emph{uniform value at $p$} if it admits an asymptotic value, and for every $\varepsilon>0$, there exists a signaling policy $\s_{\varepsilon} \in \Sigma$ and a discount factor $\delta_0$ such that $\gamma_{\delta}^x (p, \s_{\varepsilon}) \geq v_{\infty}^x(p) - \varepsilon$ for every $\delta > \delta_0$. In such a case $v_{\infty}^x (p)$ is said to be the \emph{uniform value at} $p$ of $MP(x)$.
\end{itemize}
\end{defn}

The sophistication in the
notion of the uniform value, compared to that of the asymptotic value, is that the
sender should come up with signaling policies that are \textbf{robust to changes in the
discount factor}. Formally, for any error term $\varepsilon$, the sender should come up with a policy that is $\varepsilon$-optimal  for any sufficiently large discount factor. 

Our first main result establishes the existence of the uniform value of $MP(x)$, whenever the revelation rate is positive (i.e., $x>0$), and provides a formula for it. 

As preparation for the statement of the result and the discussed formula, let us denote by $C_1,...,C_r \subseteq K$ the disjoint communication classes\footnote{A \textit{communication class} is a set $C\subseteq K$ so that for every $i,j \in C$ there is a positive probability to reach $j$ from $i$ in a finite number of steps under the transition rule $M$ and vice versa. A communication class $C$ with only one element consists of an absorbing state $i$, i.e., once $i$ is reached the Markov chain cannot exit that state.} of $M$. For every such class $C_i$, let $\pi_{C_i} \in \Delta(K)$ denote the unique stationary distribution of $M$, that is supported on $C_i$. The distribution $\pi_{C_i}$ corresponds to the unique stationary distribution within the communicating class $C_i$. Let $T = K \setminus (C_1 \cup ... \cup C_r)$. For every state $j \in T$, and $i=1,...,r$ denote by $M[j \to C_i]$ the probability to visit $C_i$ starting from state $j$, under the Markov transition matrix $M$. Note that for every $j \in T$ it holds that $\sum_{i=1}^r M[j \to C_i] = 1$. Lastly, for every $\l \in K$, the $\l$th row of $M$ is denoted by ${\textbf{m}}_{\l}$. Note that ${\textbf{m}}_{\l} = \delta_{\l}M$, so that it describes the distribution of tomorrow's state, given that the Markov chain today's state is located at $\l$.

\begin{theorem}\label{Thm1}
For every revelation rate $x \in (0,1]$ there exists a number $\V(p,x)$, given by the formula,
\begin{equation*}
\V(p,x) =  \sum_{i=1}^r \left[ p(C_i) + \sum_{j \in T} p^j \cdot M[j \to C_i] \right] \left( \sum_{\l \in C_i} \pi_{C_i}^{\l} \cdot v_{1-x}(\emph{\textbf{m}}_{\l})   \right), 
\end{equation*}
such that $v_{\delta}^x (\cdot)$ converges uniformly on $\Delta(K)$ to $\V(\cdot ,x)$ as $\delta \to 1^- $. Moreover, there exists a signaling policy $\s^* \in \Sigma$ such that for every prior $p \in \Delta(K)$, and every $\varepsilon>0$ there exists $\delta_{0} \in (0,1)$ so that 
\begin{equation*}
   \gamma_{\delta}^x (p, \s^*) \geq  \V(p, x) - \varepsilon, \quad \forall  \delta > \delta_{0}.
\end{equation*}
In particular, $\V(p,x)$ is the uniform value at $p$ for every $p \in \Delta(K)$.
\end{theorem}
\bigskip

We note that the strategy $\s^* \in \Sigma$ in Theorem \ref{Thm1} achieves the uniform value in a much stronger sense than that required in Definition \ref{Definition Uniform Value}. Indeed, it is both independent of the prior $p \in \dk$, and also does not depend on the error term $\varepsilon>0$. Put differently, a sufficiently patient sender does not need to know the initial belief $p$, nor the exact error term $\varepsilon$, as by following $\s^*$, his payoff will approximate $\V(p, x)$ up to the error term $\varepsilon$. 
\bigskip

A case of particular interest is that where $M$ is irreducible, i.e., has one communication class and therefore a unique stationary distribution, denoted $\pi_M$.  Therefore, in such a setup, we obtain that   
$$\V(p,x) =  \pi_{M}^{1} \cdot v_{1-x}(\textbf{m}_{1}) + ... + \pi_{M}^{k} \cdot v_{1-x}(\textbf{m}_{k}), \quad \forall p \in \Delta(K),$$
This observation, together with Theorem \ref{Thm1}, gives the following Corollary.

\begin{corollary}\label{Corol Thm 1}
Assume that $M$ is irreducible. Then, for every $p \in \dk$ the uniform value at $p$ of $MP(x)$ exists, and equals $$\pi_{M}^{1} \cdot v_{1-x}(\emph{\textbf{m}}_{1}) + ... + \pi_{M}^{k} \cdot v_{1-x}(\emph{\textbf{m}}_{k}).$$
\end{corollary}


\subsubsection{The effect of the revelation rate on the value}\label{Subsec Rev Effect}

In light of Theorem \ref{Thm1}, one might naturally inquire about the impact of the revelations on the sender's payoff. To address this question, we turn to several key findings from Lehrer and Shaiderman \cite{MP}.

First, it is proven in \cite{MP} that whenever $M$ is irreducible and aperiodic, the values $v_{\delta}(\cdot)$ converge uniformly (on $\dk$) towards a number $v_{\infty} \in \mathbb{R}$, which cannot surpass  $(\text{Cav}\, u)(\pi_M)$.\footnote{In terms of the current model, such a result shows the existence of an asymptotic value for $x=0$ and an irreducible and aperiodic $M$. In terms of proof techniques, the proof of such a result is carried out by an MDP reformulation for whom one can apply a Uniform Tauberian Theorem, similarly to the starting point of the proof of Theorem \ref{Thm1} (see Subsection \ref{subsec:heuristics 1} for details). The proof is then followed by a standard analysis of the recursive formula for the $N$-Ces\'aro values of the MDP, utilizing the fact that for an irreducible and aperiodic $M$, $pM^n \to \pi_M$ for every $p \in \dk$. In this regard, the proof of Theorem \ref{Thm1} is of higher complexity, requiring additional tools from Probability Theory (see Subsection \ref{subsec:heuristics 1} for details).} Second, it is shown that in specific scenarios, the equality $v_{\infty} = (\text{Cav}\, u)(\p)$ is realized. In such cases, the asymptotic value achieves its upper bound, and whenever this situation arises, the discounted value $v_{\delta}(\p)$ 
equals to $(\text{Cav}\, u)(\p)$ for every $\delta \in [0,1)$ as well.

With these results in mind, we obtain that whenever $M$ is irreducible and aperiodic and $v_{\infty} = (\text{Cav}\, u)(\p)$ it holds, 
\begin{multline}\label{Eq.Effect}
\V(p,x)  =  \p^1 v_{1-x} (\textbf{m}_1) + \cdots + \p^k v_{1-x} (\textbf{m}_k)\\ \leq v_{1-x} \left( \pi_M^1 \cdot \textbf{m}_1 + \cdots  + \pi_M^k \cdot \textbf{m}_k \right) = v_{1-x} (\p) = (\text{Cav}\, u)(\p) = v_{\infty},
 \end{multline}
for every $x \in (0,1]$ (the 
inequality follows from the concavity of $v_{1-x}(\cdot)$\footnote{For a proof of the concavity of $v_{1-x}(\cdot)$, we refer the reader to the proof of item (i) of 
Proposition 4 in Lehrer and Shaiderman \cite{MP}.}).  
When the function $v_{1-x}(\cdot)$ is not an affine one, the inequality in Eq.\ (\ref{Eq.Effect}) is sharp.  
In this case, the asymptotic value with revelations, $\V(p,x)$, is strictly lower 
than $v_{\infty}$, implying that the revelations actually
reduce the payoff of a sufficiently patient sender, regardless of how small the revelation rate is.

\bigskip

\subsection{Monotonic trajectories in the discount factor}

The values of discounted or finite dynamic games with a lack of information on one side (such as those analyzed by Aumann and Maschler \cite{Aumann}, Renault \cite{Renault}, and Lehrer and Shaiderman \cite{MP}) may vary based on the exact choice of discount factor or the length of the game, respectively. The current work introduces a new parameter: the revelation rate $x$, at which the uninformed player obtains extra information about the realized state. As it turns out, the addition of such a parameter may be utilized to obtain new insight into the analytic behavior of the values of discounted games without revelations, as a function of the discount factor. In particular, expanding the scope by allowing stochastic revelations enables one to learn more about traditional games in which such revelations are absent.

In the Aumann and Maschler \cite{Aumann} zero-sum games, when the state is selected once and for all, the value of the discounted game decreases with the discount rate. That is, the larger the discount factor, the smaller the value. The intuition is clear: a greater discount factor enables the uninformed player (the minimizer) to learn more to her advantage.

When the state is dynamic, however, the value no longer needs to be monotone with respect to the discount factor. Nevertheless, the introduction of stochastic revelations gives rise to several monotonicity results regarding the values of Markovian persuasion games: this time not about a specific prior, but rather about a combination of values at certain sets of priors. To the best of our knowledge, these are the first
monotonicity results of a similar sort.

For the sake of elegance and simplicity of the statements of our next results, we shall assume that $M$ is irreducible, and recall that $\pi_M$ denotes its unique stationary distribution. The next theorem states that a convex combination of the values $\{ v_{\d}({\textbf{m}}_{\l})\,:\, \l\in K \}$, is monotonically non-decreasing with the discount factor. 

\begin{theorem}\label{Thm Mono 1}
The mapping $$
\delta \mapsto \pi_{M}^{1} \cdot v_{\delta}(\emph{\textbf{m}}_{1}) + ... + \pi_{M}^{k} \cdot v_{\delta}(\emph{\textbf{m}}_{k})$$ is non-decreasing on $[0,1)$.
\end{theorem}
\bigskip

As it turns out, a suitable generalization of the model setup leading to Theorems \ref{Thm1} and \ref{Thm Mono 1}, may be exploited to obtain a family of monotonic trajectories, adding to the single monotonic trajectory given in Theorem \ref{Thm Mono 1}. In this regard, we learn that beneficence from patience is not an exclusive feature of the beliefs $\{ \textbf{m}_{\l}\,:\, \l \in K\}$, but rather is a global phenomenon occurring at infinitely many sets of beliefs. Such a result reads as follows.

\begin{theorem}\label{Thm Mono 2}
    Assume that the beliefs $\{\chi_1,...,\chi_k\} \subseteq \Delta(K)$ satisfy 
    \begin{itemize}
        \item[\emph{(a)}]  $\chi_1,...,\chi_k$ are affinely independent.
        \item[\emph{(b)}] $M(\Delta(K)) \subseteq conv\, \{\chi_1,...,\chi_k\}$, i.e., for every $p \in \Delta(K)$, $pM \in conv\, \{\chi_1,...,\chi_k\}$.
    \end{itemize}
    Then, the mapping $$\delta \mapsto \gamma_1 v_{\delta}(\chi_1) + ... + \gamma_k v_{\delta}(\chi_k),$$ is non-decreasing on $[0,1)$, where $\gamma_1,...,\gamma_k$ are the unique convex weights satisfying
    \begin{equation*}
        \pi_M = \gamma_1 \chi_1 + ... + \gamma_k  \chi_k.
    \end{equation*}
\end{theorem}

Condition (a), coupled with the fact that we deal with $k$ beliefs, implies that for every point $p \in conv\, \{\chi_1,...,\chi_k\}$ there exist \textbf{unique} convex weights $\g_1 (p),..., \g_k(p)$ such that $p = \gamma_1(p)\cdot  \chi_1 + ... + \gamma_k(p)\cdot \chi_k$ (where we set $\gamma_{\l} := \g_{\l}(\p)$). The quintessential property that follows from the uniqueness of $\g_1 (\cdot),..., \g_k(\cdot)$, is that $\g_{\l}(\cdot)$, viewed as a function from $M(\dk)$ to $[0,1]$ is an affine function for every $\l \in K$.

Condition (b) in the above Theorem is equivalent to the assertion that ${\textbf{m}}_{1},...,{\textbf{m}}_{k} \in conv\, \{\chi_1,...,\chi_k\}$ (as $M(\Delta(K))$ is equal to the convex hull of $\{ {\textbf{m}}_{\l}\,:\, \l\in K \}$). In geometric terms, condition (a) requires the polytope spanned by the beliefs $\chi_1,...,\chi_k$ to have a non-empty interior, i.e., to be of full dimension within $\dk$, whereas condition (b) ensures that such a polytope includes the belief $\p$.

Theorem \ref{Thm Mono 2} recovers the result of Theorem \ref{Thm Mono 1} in the case where  ${\textbf{m}}_{1},...,{\textbf{m}}_{k}$ are affinely independent, as in this case, the choice $\chi_{\l} = {\textbf{m}}_{\l}$ satisfies both conditions (a) and (b). The reader may wonder as to why the phenomenon described in Theorem \ref{Thm Mono 1} holds even when ${\textbf{m}}_{1},...,{\textbf{m}}_{k}$ are not required to be affinely independent. As we shall see in the proof of Theorem \ref{Thm Mono 1}, this is due to the special relation ${\textbf{m}}_{\l} = \d_{\l}M$, which allows us to decompose each belief of the form $pM$, where $p \in \dk$, into $pM = \sum_{\l \in K}p^{\l} \cdot \d_{\l}M  = \sum_{\l \in K}p^{\l}\cdot  {\textbf{m}}_{\l}$. For such a decomposition, the convex weights $p^1,...,p^{k}$, which are affine functions of $p$, essentially `duplicate' the role of the weights $\g_1(\cdot),...,\g_k(\cdot)$ described above.

\bigskip

The following Corollary gives a strategic interpretation for Theorems  \ref{Thm Mono 1} and \ref{Thm Mono 2}. 

\begin{corollary}\label{Corol Mono 1 and 2}
In a Markovian persuasion game (without stochastic revelations) for any two discount factors $0\leq \d<\d'<1$, there exist infinitely many prior beliefs $p \in \dk$ such that  
\begin{align*}
    v_{\d}(p) \leq v_{ \d^{'} }(p).
\end{align*}
In particular, for any two discount factors $\d<\d'$, there are infinitely many priors $p \in \Delta(K)$, from which the sender would (weakly) benefit from patience. 
\end{corollary}

The heuristics behind the proofs of the monotonicity results given in Theorems \ref{Thm Mono 1} and \ref{Thm Mono 2} are detailed in Subsections \ref{Subsec Mono Heuristics 1} and \ref{Subsec Mono Heuristics 2}, respectively.

In the next example we showcase the phenomenon described in Theorem \ref{Thm Mono 2} for a binary Markov chain. The goal of the example is to illustrate a situation where the corresponding average of values is strictly increasing, although the values themselves need not be monotone in the discount factor.

\begin{example}
Consider the case where $|K|=2$ and the matrix $M$ is given by
\begin{equation*} 
M=\begin{pmatrix}
  \ 1/2& 1/2\ \\
 \ 1/6 & 5/6\
\end{pmatrix}.
\end{equation*}
It holds that $\pi_M = (0.25,0.75)$. Also, let $u((p,1-p)) := p(2-3|p-\frac{1}{2}|)+(1-p)p/10$ where $p \in [0,1]$. It was shown in an earlier working paper version\footnote{See Example 4 on p.\ 17-18 in \href{https://www.math.tau.ac.il/~lehrer/Papers/Markovian\%20Persuasion.pdf}{https://www.math.tau.ac.il/~lehrer/Papers/Markovian\%20Persuasion.pdf}} of Lehrer and Shaiderman \emph{\cite{MP}} that
\begin{equation*}
v_{\d}((p,1-p)) = \left\{
       \begin{array}{ll}
        2.05\left(3(\d-1)p - \d/2\right)/(\d -3) , & \hbox{for\,\,}\,\, p \in [0,0.5],\\ 
        (1-\d)u((p,1-p)) + \d\cdot 2.05\left((\d-1)p - 1/2\right)/(\d -3) , & \hbox{for\,\,}\,\, p \in [1/2,1].
       \end{array}
     \right.
\end{equation*}
Let us take $\chi_1 = (0.1,0.9)$ and $\chi_2 = (0.9,0.1)$, so that $\gamma_1 = 0.8125$ and $\gamma_2 = 0.1875$. In Figure \emph{\ref{figure:Figure}} we plot the $\d$-trajectories of the values at $\chi_1$ and $\chi_2$, together with the trajectory of their corresponding average. In the figure we abuse notation and denote $v_{\d}(p,1-p)$ by $v_{\d}(p)$. As depicted in the figure, the trajectory $\d \mapsto v_{\d}(\chi_2)$ is a non-monotone function, whereas $\d \mapsto \gamma_1 v_{\d}(\chi_1)+\gamma_1 v_{\d}(\chi_2)$ is strictly increasing.

\begin{figure}[htp] \centering{
\includegraphics[scale=0.33]{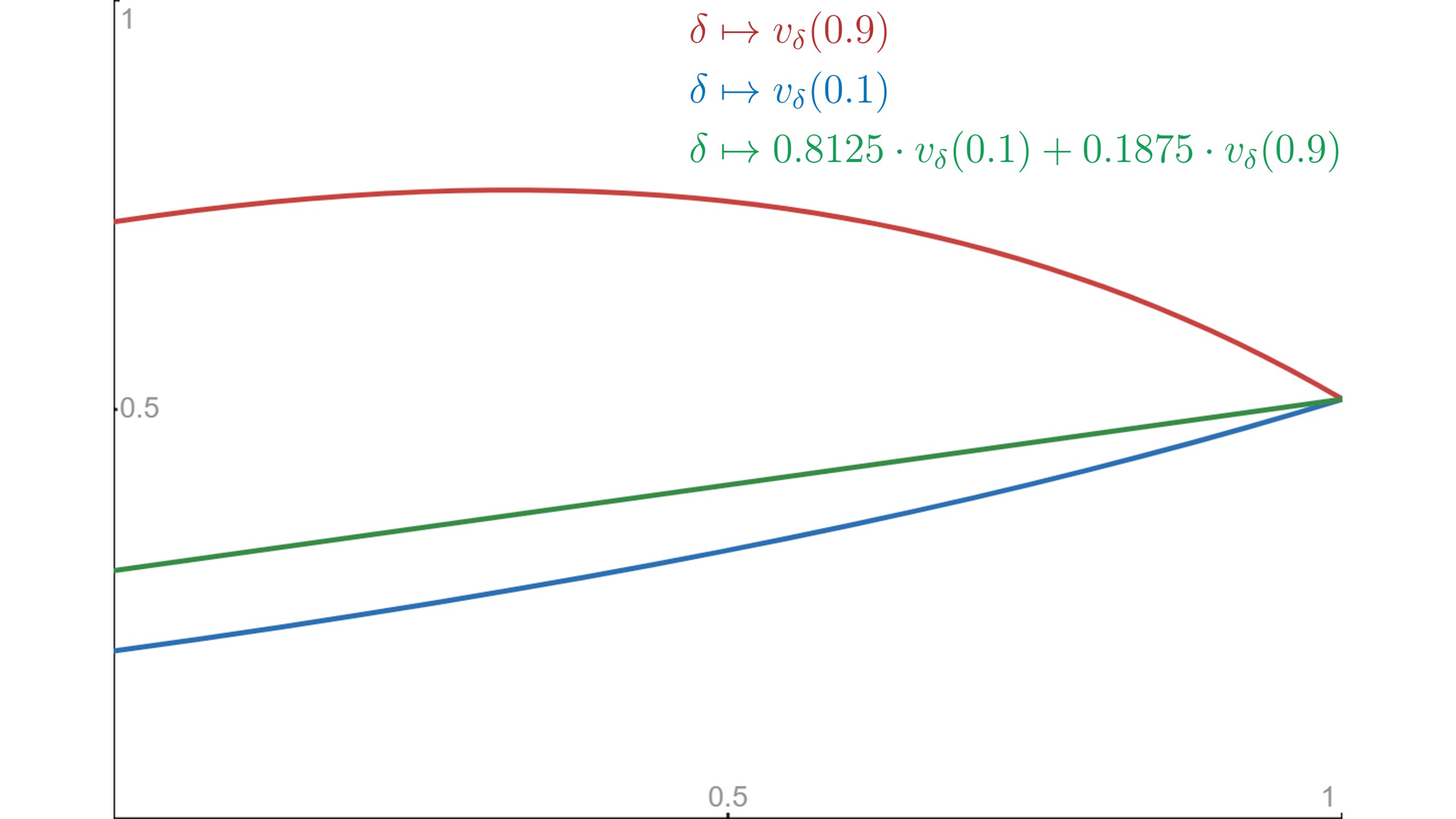}}
\caption{$\delta$-Trajectories of the values}
\label{figure:Figure}
\end{figure}
\end{example}

To end the discussion of the main results, we encourage the readers of the paper to seek to understand whether the proof techniques of the current paper have the potential to be useful in other models as well (see also Section \ref{sec: application to repeated games}). Albeit subjective, it is our hope and belief that the introduction of suitable exogenous parameters to dynamic games may give additional insight into the baseline models of such games. 

\section{The heuristics behind the main results}\label{sec:heuristics}

The current section is devoted to the introduction of the main ideas and techniques used for the proofs of the main results. We encourage the reader to read this section prior to delving into the formal proofs, as it was intended to smooth the transition into formalities. Lastly, we note that this section is written in chronological order, in the sense that each subsection relies in some capacity on the ideas presented in the subsections preceding it.  

\subsection{The heuristics behind the proof of Theorem \ref{Thm1}}\label{subsec:heuristics 1}

\headings{A general MDP approach} 

 The first step towards proving Theorem \ref{Thm1} is to embed the game $MP(x)$ in a more general Markov Decision Process (MDP) framework. Employing a Tauberian Theorem for MDPs due to Lehrer and Sorin \cite{Lehrer}, we reduce the problem of proving that $v_{\delta}^x (\cdot)$ converges uniformly to $\V(\cdot, x)$, to  proving that $v_N^x(\cdot)$ converges uniformly to $\V(\cdot, x)$, where $v_N^x(\cdot)$ is the value with respect to the $N$-Ces\`aro average payoff, formally defined by,
\begin{equation*}
v_N^x(p) := \max_{\s \in \Sigma} \left[ E^p_{x,\s} \left(\frac{1}{N}\sum_{n=1}^N u(p_n)\right) \right], \quad  \forall p \in \dk.
\end{equation*}
The discussed MDP is introduced, motivated, and surveyed in Section \ref{Section General MDP}. 

\headings{Informational reboots and revelations.}

 By the Markov property of $(X_n)_{n \geq 1}$, upon a revelation of the realized state of a step in $(X_n)_{n \geq 1}$, the information obtained by the receiver from all previous signals perishes. For instance, if the realized state at the first time a revelation occurred was $\l \in K$, the prior belief of the receiver about tomorrow's state is ${\textbf{m}}_{\l} = \d_{\l}M$, regardless of past signals observed or the stage at which the revelation
occurred. In view of this, one can think of the revelations as reboots: after each reboot, there is a renewal of information. 


 Such an informational phenomenon has important strategic implications. In fact, on stages following revelations, the game starts anew with a prior belonging to $\{ {\textbf{m}}_{\l}\,:\, \l \in K\}$. In particular, starting from such a stage, and in the stages that follow until the next revelation, the sender and receiver engage in a Markovian persuasion game with no revelations. The $N$-Ces\`aro average payoff comes into play, as it incentivizes the sender to maximize his \textbf{expected sum of payoffs} along the random-duration Markovian persuasion games occurring between any two successive revelations.

\headings{The random-duration valuation}

 The previous observation gives rise to the notion of a \textit{random-duration} valuation (see Subsection \ref{Subsection Random Duration Valuation}). In such a valuation, the total payoff is the expected sum of stage payoffs in a random number of stages, which is distributed geometrically. Such a valuation applies to our setup, as the number of stages between any two successive revelations follows a geometric distribution with the revelation rate $x$ as its parameter. 

As it turns out, the value of a Markovian persuasion game under such a valuation admits a representation in terms of the $\d$-discounted value. In detail, for any prior $p \in \dk$, the value of a Markovian persuasion game under a random-duration valuation having $x$ as its geometric parameter equals $v_{1-x}(p)/x$ (as follows from Lemma \ref{Random Duration Value Lemma}). The underlying rationale behind this phenomenon is rooted in the fact that a $\d$-discounted payoff awards the sender with his payoff on the \textbf{last stage of play only} (rather than the sum of stage payoffs), where the total number of stages played is distributed geometrically with parameter $\d$. 

The random-duration valuation and its properties are introduced in Subsection \ref{Subsection Random Duration Valuation}, in the broader MDP framework.  

\headings{The optimal strategy $\s^*$}

 The informational reboots together with the random-duration
valuation, inspire the strategy $\s^*$, described by:
\begin{itemize}
    \item Play arbitrarily until the first revelation.
    \item Given that $X_n = \l$ was revealed, starting from stage $n+1$, play optimally in a Markovian persuasion game with random duration valuation, having ${\textbf{m}}_{\l}$ as its prior, until the next stage in which a revelation occurs (including that stage). 
\end{itemize}
The strategy $\s^*$ is independent of the initial prior $p$. Moreover, in essence,  it `breaks' the game $MP(x)$ into so-called `random-duration games', in each of which it plays myopically. In Subsection \ref{Subsec nu-star} we define and examine the natural generalization of $\s^*$ to the MDP framework.

\headings{The main technical challenges}

The $N$-Ces\`aro average payoff admits the following challenge. Assume that the total number of revelations up to stage $N$ was known. Then, by following $\s^*$, the sender could have guaranteed a corresponding average of $v_{1-x}({\textbf{m}}_{\l})/x$, based on the proportion of revelations at which state $\l$ was the revealed state. However, as the total number of revelations up to stage $N$ is \textbf{random}, at face value, it is not clear how one can \textbf{extract} the values $v_{1-x}({\textbf{m}}_{\l})/x$ of the `random-duration games' from the \textbf{expected} $N$-Ces\`aro average payoff under $\s^*$, let alone understand how to average them properly. 

\headings{Random times and the Ergodic Theorem}

To deal with the mentioned challenges, we introduce in Subsection \ref{Subsec Prob Approach} a sequence of \textbf{random times} $T_1,...,T_m,...$ within the MDP framework, which in the current setup correspond to the sequence of stages at which revelations occur. Namely, the $m$th revelation occurs at stage $T_m$. For such a sequence, we study an auxiliary sequence of payoffs $(\mathcal{Q}_m)_{m\geq 1}$, which for the Markovian persuasion model can be written as,
\begin{align*}
    \mathcal{Q}_m(p,\s) & = E^p_{x,\s} \left(\sum_{n=T_1+1}^{T_{m}} u(p_n)\right)\\
    &= \sum_{i=1}^{m-1} E^p_{x,\s} \left(\sum_{n=T_i+1}^{T_{i+1}} u(p_n)\right), \quad  \forall p \in \dk, \, \forall \s \in \S.
\end{align*}
Thus, $\mathcal{Q}_m$ captures the sum of expected payoffs along the first $m-1$ random-duration games occurring in between the first $m$ revelations. 

The definition of $\s^*$ suggests that it should be optimal under the payoff $\mathcal{Q}_m$ for every $m\geq 1$, i.e.,  $\mathcal{Q}_m(\cdot ,\s)\leq \mathcal{Q}_m(\cdot ,\s^*)$ for any $\s\in \Sigma$. The corresponding optimality in the broader MDP framework is addressed in Subsection \ref{Subsec nu-star optimality}. To link the payoffs $(\mathcal{Q}_m)_{m\geq 1}$ to the $N$-Ces\`aro average payoffs, we employ in Subsection \ref{Subsec Prob Approach} probabilistic arguments for the MDP analysis, which subsequently imply that for any $p \in \dk$ and $\s \in \S$:
\begin{align}\label{Eq. Approx Heuristics}
    \left\vert \frac{1}{L_m} \mathcal{Q}_m(p,\s) -  E^p_{x,\s} \left(\frac{1}{L_m}\sum_{n=1}^{L_m} u(p_n)\right) \right\vert \leq o_m(1),
\end{align}
where $L_m = \lfloor E^p_{x,\s}\, T_m \rfloor = \lfloor m/x \rfloor$, and $o_m(1)$ denotes a function of $m$ and $x$ which vanishes as $m\to \infty$. Such asymptotic proximity, together with the optimality of $\s^*$ under $\mathcal{Q}_m$, implies that 
\begin{align}\label{Eq. Heuristics}
    \left\vert \frac{1}{L_m} \mathcal{Q}_m(p,\s^*) -  v_{L_m}(p) \right\vert \leq o_m(1).
\end{align}

As the sequence $(L_m)_{m\geq 1}$ has bounded increments (i.e., $L_{m+1}-L_m \leq 1/x$), the uniform convergence of $\{v_{L_m}(p): m\geq 1\}$ is sufficient for that of $\{v_{N}(p): N\geq 1\}$. Thus in view of Eq.\ \eqref{Eq. Heuristics}, it suffices to show the uniform convergence of $\mathcal{Q}_m(\cdot ,\s^*)/L_m$ to $\V(\cdot,x)$ on $\dk$. Let us now sketch the key steps leading to such convergence. First, the definition of $\s^*$ and $\mathcal{Q}_m$, coupled with our random-duration valuation discussion, would imply that
\begin{align*}
    \frac{1}{L_m} \mathcal{Q}_m(p,\s^*) & = \sum_{\l \in K} \left( \frac{1}{L_m} \sum_{i=1}^{m-1} P^p_{x,\s} \left(X_{T_i} = \l\right) \right)\cdot v_{1-x}({\textbf{m}}_{\l})/x\\
    & = \sum_{\l \in K} \left( \frac{1}{L_m} \sum_{n=1}^{L_m} P^p_{x,\s} \left(X_{n} = \l, Z_n =1 \right) \right)\cdot v_{1-x}({\textbf{m}}_{\l})/x + o_m(1),
\end{align*}
where the approximation in the second equality is a particular case of a more general approximation within the MDP framework showcased in Subsection \ref{Subsec nu-star payoff}. Next, as one has that
\begin{align*}
    P^p_{x,\s} \left(X_{n} = \l, Z_n =1 \right) & = x \cdot P^p_{x,\s} (X_n = \l )\\
    & = x\cdot \left(pM^{n-1} \right)^{\l}\\
    & = x\cdot \sum_{j \in K} p^j \cdot \left(\d_j M^{n-1} \right)^{\l}
\end{align*}
we obtain using the preceding approximation that 
\begin{align}\label{Eq. Approx Heuristics 2}
    \frac{1}{L_m} \mathcal{Q}_m(p,\s^*) &   = \sum_{\l \in K} \left(\, \sum_{j \in K} p^j \left[ \frac{1}{L_m} \sum_{n=1}^{L_m} \left(\d_j M^{n-1} \right)^{\l} \right]  \right) \cdot v_{1-x}({\textbf{m}}_{\l}) + o_m(1).
\end{align}
Thus, it remains to assess the asymptotic behavior of $(1/L_m)\sum_{n=1}^{L_m} \left(\d_j M^{n-1} \right)^{\l}$. In particular, we note that $(1/L_m)\sum_{n=1}^{L_m} \left(\d_j M^{n-1} \right)^{\l}$ describes the \textbf{expected} frequency of visits to state $\l$ by $X_1,...,X_{L_m}$ starting from $X_1 =j$. Such an assessment is parsed based on the classification of $\l$. If $\l \in T$, then its limiting frequency of visits is $0$. Otherwise, if $\l$ is in the communication class $C_i$, the Ergodic Theorem for irreducible Markov Chains (e.g., Theorem C.1 in \cite{Peres}) implies that 
\begin{itemize}
    \item Starting from any state $j \in C_i$, the frequency of visits of $\l$ tends (almost surely) to $\pi^{\l}_{C_i}$, implying that $(1/L_m)\sum_{n=1}^{L_m} \left(\d_j M^{n-1} \right)^{\l} \to \pi^{\l}_{C_i}$ as $m \to \infty$.
    \item Starting from any $j \in C_m$, where $m\neq i$, there is a zero probability of reaching $\l$, and thus the frequency of visits of $\l$ vanishes.
    \item Starting from $j \in T$, the Markov chain is absorbed (in finite time) into $C_i$ with probability $M[j\to C_i]$. Thereafter, the limiting frequency equals (almost surely) $\pi^{\l}_{C_i}$ as before.
\end{itemize}
A combination of the above arguments with the approximation in \eqref{Eq. Approx Heuristics 2} implies that 
\begin{align*}
    \frac{1}{L_m} \mathcal{Q}_m(p,\s^*) &   =  \sum_{i=1}^r \left[ p(C_i) + \sum_{j \in T} p^j \cdot M[j \to C_i] \right] \left( \sum_{\l \in C_i} \pi_{C_i}^{\l} \cdot v_{1-x}(\emph{\textbf{m}}_{\l})   \right) + o_m(1),
\end{align*}
which provides the desired formula for the uniform value stated in Theorem \ref{Thm1}. Formally, the proof of Theorem \ref{Thm1} is a result of a generalized result for the MDP framework, stated in Theorem \ref{Uniform Value for MDP} in Subsection \ref{Subsec MDP plan}. In essence, the proof of such a generalized result, given in Section \ref{Proof of Uniform and Asym Values}, is achieved by formalizing the above-mentioned heuristics to the MDP framework.

\subsection{The heuristics behind the proofs of Theorem \ref{Thm Mono 1}}\label{Subsec Mono Heuristics 1}

\headings{The sufficient condition - monotonicity of $v_N^x$ in $x$}

In the discussion detailing the heuristics behind the proof of Theorem \ref{Thm1}, we introduced the values $v_N^x(\cdot)$ with respect to the $N$-Ces\`aro average payoff, which, in fact, converge uniformly to $\V(\cdot,x)$. Let us now argue that in order to prove Theorem \ref{Thm Mono 1}, it suffices to show that for any $N\geq 1$, and any prior $p \in \dk$, the mapping $x \mapsto v_N^x(p)$ is non-increasing on $(0,1]$. Indeed, in the case where $M$ is irreducible, by Theorem \ref{Thm1}, the mapping $x \mapsto \pi_{M}^{1} \cdot v_{1-x}(\textbf{m}_{1}) + ... + \pi_{M}^{k} \cdot v_{1-x}(\textbf{m}_{k})$ must also be non-increasing in $x$, being the point-wise limit of the sequence of non-increasing mappings $\{x \mapsto v_N^x(p)\,:\, N\geq 1\}$. This in turn with the fact that $\d \mapsto 1-\d$ is non-increasing on $[0,1)$, establishes that $\d \mapsto \pi_{M}^{1} \cdot v_{\d}(\textbf{m}_{1}) + ... + \pi_{M}^{k} \cdot v_{\d}(\textbf{m}_{k})$ must be non-decreasing, as desired. 

\headings{The underlying coupling by the sender}

The main heuristic argument behind the monotonicity of $x \mapsto v_N^x(p)$ is rooted in a probabilistic coupling of two Bernoulli random variables. Take $0<x<y\leq 1$. Then, in the game $MP(x)$, at the start of each period $2\leq n\leq N$, whenever $Z_{n-1}=0$, the sender \textbf{`can tell'} the realized state $x_{n-1}$ to the receiver with probability $(1-y)/(1-x)$. By doing so, at the start of each stage  $2\leq n\leq N$, the receiver will be aware of the realized state of $X_{n-1}$ with probability $x + (1-x)\times (1-y)/(1-x) =y$, just as in $MP(y)$. In essence, this means that the game $MP(y)$ may be embedded in $MP(x)$, and therefore its $N$-Ces\`aro values $\{v_N^y\,:\, N\geq 1\}$ cannot surpass those of $MP(x)$. 

\headings{The technical challenges}

The formalization of the coupling argument encounters two technical challenges. The first arises from the finiteness of the signal set $S$; can the sender both tell the preceding realized state and transmit an informative signal about the current realized state using a single signal $s \in S$? If $S$ is of cardinality $k$, clearly not. 

The second challenge arises from the fact that the coupling hands out the revelation in a single-stage lag. Although somewhat vague at this point, such a lag is of importance, as the \textit{`prior belief'} of the receiver at stage $n$, being the $M$-shift, $p_{n-1}M$, of the preceding posterior $p_{n-1}$, is not influenced by the coupling, which may be carried out later only using the signal $s_n$. The latter will be of consequence,  
requiring the sender to make his coupling strategic, in the sense that it would also have to account for the prior belief $p_{n-1}M$, and not just the value of $Z_{n-1}$.

\headings{Recursive formulas and Carath\'{e}odory Theorem }

The main tool used to tackle both of the above challenges is taken from a dynamic-programming principle, which allows for recursive formulas for $v_N^x$, for any $N\geq 1$. Formally introduced in Subsection \ref{Subsubsec Recursive Formula}, such  
formulas have two important derivatives.

First, their combination with Carath\'{e}odory Theorem from convex analysis (see, e.g., Corollary 17.1.5 in \cite{Rock}), shows the invariance of $v_N^x$ to the cardinality of $S$, as long as the latter exceeds $k$. This resolves the first challenge, as it allows us to assume that the sender has an infinite number of signals at his disposal. Second, those formulas play a key role in overcoming the 
`single-stage lag challenge', as they allow us to concentrate on signaling strategies which only take into account at each stage $n=1,...,N$, the prior belief $p_{n-1}M$ together with the value of $n$. The formalization of the above two derivatives can be found in Subsection \ref{Subsubsec Optimal Strategies}.

Lastly, the exact coupling and its validity appear in Subsection \ref{Subsection Monotonicity Proof}, which is devoted to the formal proof of the monotonicity of the mappings $\{x \mapsto v_N^x(p)\,:\, N\geq 1\}$.

\subsection{The heuristics behind the proof of Theorem \ref{Thm Mono 2}}\label{Subsec Mono Heuristics 2}

\headings{A  model with partial stochastic revelations}

The proof of Theorem \ref{Thm Mono 2} is achieved by means of a suitable modification of the model setup studied in the paper. Namely, one can think of the stochastic revelations as being \textit{Blackwell experiments} or, equivalently \textit{splits} (see e.g., Blackwell \cite{Blackwell} and Aumann and Maschler \cite{Aumann}). At the end of each stage $n\geq 1$, the posterior $p_n$ is split to the extreme points $\{\d_{\l}\,:\, \l \in K\}$ of $\dk$.

Consider now a scenario where instead of the above setup, starting from the second stage, with probability $x$, independently of all past and future plays, the prior at stage $n$, $q_n = p_{n-1}M$, is split to the beliefs $\chi_1,...,\chi_k$ according to the convex weights $\g_1 (q_n),...,\g_k(q_n)$. In detail, at the start of each stage $n\geq 2$, the receiver and the sender observe the Bernoulli trial $Z_n$, which, in the event of success, gives the receiver the additional (partial) information described by the mentioned split. Only thereafter does the sender observe the realized value $x_n$ of $X_n$, and the timeline for the remainder of the stage consists of the strategic decisions made by both players.

\headings{Repeating the machinery}

The analysis of the above-modified model builds on the same techniques and arguments required for the proofs of Theorems \ref{Thm1} and \ref{Thm Mono 1}. Based on parallel proof heuristics to those of Theorem \ref{Thm1}, it is shown formally in Subsection \ref{Proof of Mono 2 Convergence}, that $v_N^x(\pi_M) \to \gamma_1 v_{1-x}(\chi_1) + ... + \gamma_k v_{1-x}(\chi_k)$ as $N \to \infty$.

Thereafter, generalizing the coupling technique in the proof of Theorem \ref{Thm Mono 1}, it is shown in Subsection \ref{Proof of Mono 2 Monotonicity} that the mappings $\{x \mapsto v_N^x(\pi_M)\,:\, N\geq 1\}$ should be non-decreasing in $x$, just as in the baseline model. In essence, such coupling is based on the following argument: for $0<x<y\leq 1$, when playing the game with revelation rate $x$, at any stage $n=2,...,N$, in case $Z_n =0$, the sender can split the prior belief $p_{n-1}M$ to $\chi_1,...,\chi_k$ with probability $(y-x)/(1-x)$. 

As in the proof of Theorem \ref{Thm Mono 1}, the combination of the two properties above yields the result of Theorem \ref{Thm Mono 2}.

\section{Application to zero-sum games with one-sided information and changing states}\label{sec: application to repeated games}

The Markovian persuasion model shares similarities with Renault's model of dynamic zero-sum games with incomplete information on one side (see \cite{Renault}). In the current paper, by allowing for stochastic revelations, we extend the baseline Markovian persuasion model. A similar extension may be considered for Renault's model.  

For the description of such an extension, we associate with each state $\l \in K$ a zero-sum matrix game $G^{\l}$. We assume that the matrices $\{G^{\l}\}_{\l \in K}$ are of equal dimensions. The action set of the row player (the maximizer) in the games $\{G^{\l}\}_{\l \in K}$ is denoted $I$, whereas that of the column player (the minimizer) by $J$. The payoff for the row player in $G^{\l}$ under the action profile $(i,j) \in I \times J$ is denoted by $G^{\l}[i,j]$. The row player and column player, referred to as Player 1 and Player 2, resp., engage in a repeated zero-sum game, denoted $RG(x)$. In that game, the processes $(X_n)_{n\geq 1}$ and $(Z_n)_{n\geq 1}$ described in Section \ref{sec: model} also serve as primitives, whose laws are known to both players.  

The game $RG(x)$ is played by stages, and the timeline of each stage $n\geq 1$ consists of the following sequence of events:
\begin{itemize}
    \item Player 1 observes the realized state $x_n \in K$ of $X_n$.
    \item Players 1 and 2 take simultaneously actions $i_n \in I$ and $j_n \in J$.
    \item The action profile $(i_n,j_n)$ is publicly announced.
    \item Player 1 receives from Player 2 the payoff $G^{x_n}[i_n,j_n]$. Note that unlike Player 1, Player 2 does not know his stage payoff, which equals $-G^{x_n}[i_n,j_n]$, as he is not informed of $x_n$.
    \item  Both players observe the outcome of the Bernoulli trial $Z_n$. The state $x_n$ is made public if and only if the trial is successful ($Z_n=1$). In this case, we say that a \emph{revelation} occurred. 
    \item The game proceeds to stage $n+1$.
\end{itemize}

As in Subsection \ref{subsection model posteriors}, the random signal $R_n$ with values in $K \cup \{\emptyset \}$  is defined by $R_n=X_n$ on $\{Z_n=1\}$ and $R_n=\emptyset$ on $\{Z_n=0\}$, and its realized value is denoted by $r_n$.

A \textit{behavioral strategy for Player 1} in $RG(x)$ is described by a sequence $\s = (\s_n)_{n\geq 1}$, where $\s_n : (K\times I \times  J \times \{ K \cup \{\emptyset \} \})^{n-1} \times K \to \Delta(I)$ for every $n\geq 1$. That is, Player 1 can draw his $n$'th stage action $i_n$ from a lottery over $I$ which may depend on his past information $x_1,i_1,j_1,r_1,...,x_{n-1},i_{n-1},j_{n-1},r_{n-1}$ together with today's information, being $x_n$. Analogously, a \textit{behavioral strategy for Player 2} in $RG(x)$ is described by a sequence $\t = (\t_n)_{n\geq 1}$, where $\t_n : (I \times  J \times \{ K \cup \{\emptyset \} \})^{n-1} \to \Delta(J)$ for every $n\geq 1$. Let $\Sigma$ and $\mathcal{T}$ denote the space of behavioral strategies of Players 1 and 2, respectively.

By similar arguments to those in Subsection \ref{subsec Model Probability space}, each pair of behavioral strategies $(\s,\t) \in \Sigma \times \mathcal{T}$, together with $(X_n)_{n\geq 1}$ and $(Z_n)_{n\geq 1}$, induce a unique probability measure on $(K\times I \times  J \times \{ K \cup \{\emptyset \} \})^{\mathbb{N}}$, denoted by $P_{x,\s,\t}^p$. Let $E_{x,\s,\t}^p$ denote the expectation operator with respect to $P_{x,\s,\t}^p$. The $\d$-discounted payoff in $RG(x)$, associated with the strategy pair $(\s,\t) \in \Sigma \times \mathcal{T}$, and the prior $p$, is denoted by $\gamma_{\d}^x (\s,\t)$, and is defined by
\begin{align*}
    \gamma_{\d}^x (\s,\t) := E_{x,\s,\t}^p \left( (1-\d) \sum_{n=1}^{\infty} \d^{n-1} G^{X_n}[i_n,j_n]\right).
\end{align*}
The $\d$-discounted value of $RG(x)$, denoted $V_{\d}^x (p)$, is the scalar
\begin{align*}
 \sup_{\s \in \Sigma}\, \inf_{\t \in \mathcal{T}} \, \gamma_{\d}^x (\s,\t)
 = \inf_{\t \in \mathcal{T}} \, \sup_{\s \in \Sigma} \, \gamma_{\d}^x (\s,\t),
\end{align*}
where the equality follows from an application of Sion's Minimax Theorem (e.g., Theorem I.1.1 on p.\ 5 in \cite{MSZ book}). As before, the shorter notation $V_{\d} (p)$ is used for $V_{\d}^0 (p)$. The value $V_{\d} (p)$ agrees with the $\d$-discounted values of the dynamic model of repeated games with incomplete information on one side considered in Renault \cite{Renault}.

The first main result concerning $RG(x)$ is analogous to that given in Theorem \ref{Thm1} for $MP(x)$. It formally reads as follows.

\begin{theorem}\label{Thm:RG-1}
For every revelation rate $x \in (0,1]$ there exists a number $V^*(p,x)$, given by the formula,
\begin{equation*}
V^*(p,x) =  \sum_{i=1}^r \left[ p(C_i) + \sum_{j \in T} p^j \cdot M[j \to C_i] \right] \left( \sum_{\l \in C_i} \pi_{C_i}^{\l} \cdot V_{1-x}(\emph{\textbf{m}}_{\l})   \right), 
\end{equation*}
such that $V_{\delta}^x (\cdot)$ converges uniformly on $\Delta(K)$ to $V^*(\cdot ,x)$ as $\delta \to 1^- $. Moreover, there exists a behavioral strategy $\s^* \in \Sigma$, such that for every prior $p \in \Delta(K)$, and every $\varepsilon>0$ there exists $\delta_{0} \in (0,1)$ so that 
\begin{equation*}
   \gamma_{\d}^x (\s^*,\t) \geq  V^*(p, x) - \varepsilon, \quad \forall \t \in \mathcal{T}, \, \forall  \delta > \delta_{0}.
\end{equation*}
\end{theorem}

Our second result regarding $RG(x)$ is in the spirit of Theorem \ref{Thm Mono 2}. However, we need to add the standard assumption of \textit{cheap-talk} in $RG(x)$, often used when dealing with discounted payoffs. This assumption asserts that at each stage $n\geq 1$, before taking their respective actions $i_n$ and $j_n$, Players 1 and 2 may communicate with each other. In such a communication, the players can repeatedly transmit to their adversary any message they desire. 

Note that this assumption does not restrict generality, and thus does not affect the values of $RG(x)$, for any $x \in [0,1]$. This is due to the fact that $RG(x)$ is a \textbf{zero-sum game}. Thus, any information provided by each of the players using `cheap-talk', can only extend the strategy space of his opponent, and thus potentially decrease his payoffs.\footnote{Although formally the `cheap-talk' assumption requires updating the definition of the behavioral strategies of the players, as all the proofs remain the same, we do not dive into the exact formalities.}

The result can now be stated as follows. 

\begin{theorem}\label{Thm RG-mono}
Assume w.l.o.g.\ that cheap-talk is allowed in $RG(x)$. Then, for an irreducible $M$ with unique stationary distribution $\pi_M \in \dk$ and any set of beliefs $\{\chi_1,...,\chi_k\} \subseteq \Delta(K)$ satisfying 
    \begin{itemize}
        \item[\emph{(a)}]  $\chi_1,...,\chi_k$ are affinely independent.
        \item[\emph{(b)}] $M(\Delta(K)) \subseteq conv\{\chi_1,...,\chi_k\}$, i.e., for every $p \in \Delta(K)$, $pM \in conv\{\chi_1,...,\chi_k\}$.
    \end{itemize}
    the mapping $$\delta \mapsto \gamma_1 V_{\delta}(\chi_1) + ... + \gamma_k V_{\delta}(\chi_k),$$ is non-decreasing on $[0,1)$, where $\gamma_1,...,\gamma_k$ are the unique convex weights satisfying
    \begin{equation*}
        \pi_M = \gamma_1 \chi_1 + ... + \gamma_k  \chi_k.
    \end{equation*}    
\end{theorem}

\section{A general MDP framework}\label{Section General MDP}

The cornerstone step for the analysis in the paper is the embedding of both the  Markovian persuasion and zero-sum games
with one-sided information and changing states, having stochastic revelations, in a general Markov decision problem (MDP) framework. Such an MDP, which is denoted by $\Gamma_x$, where $x$ is the revelation rate, is defined in the broadest way possible, to make it possibly applicable to other models of dynamic games with incomplete information.

The formal description of $\G_x$ is given in Subsection \ref{Description of MDP}, after which we provide in detail in Subsection \ref{Intuition of MDP} the intuition behind the various aspects of $\G_x$. In Subsections \ref{Classic Valuations} and \ref{Subsection Random Duration Valuation}, we introduce the strategy spaces in $\G_x$, and discuss different types of payoff valuations for $\G_x$, which are required for subsequent analysis.
The formal reductions of $\G_x$ to the Markovian persuasion and zero-sum games
with one-sided information and changing states, equipped with stochastic revelations, are given in Subsections \ref{MDP of MP(x)} and \ref{MDP of ZSRG}. With the goal of trying to make the reduction as clear as possible, we added elaborate discussions tackling the main intricacies of the reduction (see, for instance, Subsection \ref{Sub.Sec:Equivalence}).

Lastly, we remark that this section is written in chronological mathematical order, so we advise the reader to follow the different discussions in their original flow.

\subsection{The formal description of $\G_x$}\label{Description of MDP}

The \textit{states} of $\Gamma_x$ are $\dk \cup \R$, where $\R = \{\d^*_{\l}M\,:\, \l \in K \}$ is a distinguished copy of the $M$-shifted Dirac measures $\{\d_{\l}\}_{\l \in K}$. The \textit{actions} in $\Gamma_x$ are lotteries $\x$ over a countable set $A$ which may depend on $K$. Such a lottery $\x$ may be described by a vector of $k$ independent lotteries $(\x^1,...,\x^k)$ over $A$. In mathematical terms, the set of actions of $\Gamma_x$ equals $\D(A)^K$.   
The \textit{payoff} $f(\o,\x)$ associated with state $\o \in \dk \cup \R$ and action $\xi \in \D(A)^K$ is given by a positive measurable bounded function $f: \{\dk \cup \R\} \times \D(A)^K \to \mathbb{R}$. The \textit{transition rule}, denoted $t$, is defined by
\begin{equation}\label{transition rule}
t(\o,\x) = \left\{
       \begin{array}{ll}
        \d^*_{\l} M , & \hbox{with prob.\,\,}\,\, x\o^{\l},	\,\,\,\, \l \in K,\\ 
        \o(\x,a) M , & \hbox{with prob.\,\,}\,\, (1-x)\sum_{\l \in K} \o^{\l}\cdot \x^{\l}(a),	\,\,\,\, a \in A,
       \end{array}
     \right.
\end{equation}
where $\o(\x,a) \in \dk$ assigns to each state $\l \in K$ a probability mass of $\o^{\l}\cdot \x^{\l}(a)/\sum_{i \in K} \o^i \cdot \x^i (a)$.\footnote{In our description, we use the convention that for $\o = \d^*_{i}M \in  \R$, $\o^{\l}$ equals to the mass assigned by the distribution $\d_{i}M \in \dk$ to state $\l$.} 

\subsubsection{The intuition behind $\G_x$}\label{Intuition of MDP}

In simplest terms, the MDP $\G_x$ is intended to describe a learning process regarding the current state of a Markov chain $(X_n)_{n\geq 1}$ attaining values in $K = \{1,...,k\}$. One component of the learning is controlled by the actions of the decision maker who observes $(X_n)_{n\geq 1}$. Another component is governed by exogenous revelations, which fully reveal the state of an underlying Markov chain $(X_n)_{n\geq 1}$ with stochastic transition rule given by the matrix $M$.

The states of $\Gamma_x$ describe possible prior beliefs over the current state of $(X_n)_{n \geq 1}$; at any stage $n\geq 1$, the current state $\o_n$ of $\Gamma_x$ describes a prior over the distribution of $X_n$. This \textit{prior} is updated to a \textit{posterior} based on the actions of the decision maker and the exogenous stochastic revelations. The next day's state of $\Gamma_x$, $\o_{n+1}$, is achieved by applying the $M$-shift of the latter posterior, shifting it to the prior distribution over $X_{n+1}$. 

The component $\R$ of the states of $\Gamma_x$ is introduced to distinguish between revelations of $(X_n)_{n\geq 1}$ occurring as a result of exogenous revelations, compared to those governed by the decision maker. Thus, when state $\o_{n}$ moves to state $\d^*_{\l}M$ we learn that the information $X_n = \l$ was revealed exogenously. As in this case, the `learning' regarding $X_n$ is `full', the action of the decision maker at stage $n$ does not contribute any more to the learning, so that the \textit{posterior} becomes $\d^*_{\l}$. The next state, $\o_{n+1}$, is therefore set to equal $\d^*_{\l}M$. Lastly, the definition in \eqref{transition rule} implies that exogenous revelations occur at random at each possible state $\o$ of $\Gamma_x$ with probability $x$, independently of $\o$ and of the actions taken by the decision maker.

In the absence of an exogenous revelation, the action of the decision maker governs the move from a \textit{prior} belief to a \textit{posterior} belief. Denoting by $\x_n$ the action at stage $n \geq 1$, the posterior is determined by $\o_n$ and the element $a \in A$ chosen by the lottery $\x$. As contingent on $X_n=\l$, each $a \in A$ is drawn with probability $\x_n^{\l} (a)$, Bayes' law ensures that given $\o_n$ and 
$\x_n$, conditional on the outcome $a$ of $\x_n$, the posterior over $K$, denoted $\o_n(\x_n,a)$, takes the form that is described right after \eqref{transition rule}. This posterior is then shifted by $M$ to the next prior $\o_{n+1} = \o_n(\x_n,a)M$. Lastly, note that the total probability of the outcome $a$ in the absence of an exogenous revelation equals $(1-x)\sum_{\l \in K} \o_n^{\l}\cdot \x^{\l}_n(a)$, in consistency with \eqref{transition rule}.

\subsubsection{The $N$-Ces\`aro and $\delta$-discounted values of $\G_x$.}\label{Classic Valuations}

A \textit{behavioral strategy} $\nu$ in $\G_x$ is described by a sequence of stage strategies $(\nu_n)_{n\geq 1}$. At each stage $n\geq 1$, $\nu_n$ associates to each history of past states and actions $\o_1,\x_1,...,\o_{n-1},\x_{n-1}$, together with today's state $\o_n$, a mixed action over $\D(A)^K$. Let $\mathcal{N}$ denote the space of all behavioral strategies in $\G_x$. For each $\nu \in \mathcal{N}$ we denote by $P^{\o}_{x,\nu}$ the unique probability measure describing the distribution over infinite histories $(\o_1,\x_1,...,\o_n,\x_n,...)$, generated by the strategy $\nu$ and the initial state $\o_1 = \o$. Let
\begin{align*}
    \U^x_N (\o, \nu) :=  E^{\o}_{x,\nu} \left( \frac{1}{N} \sum_{n=1}^N f(\o_n,\x_n)  \right), \quad N \in \mathbb{N},
\end{align*}
and 
\begin{align*}
    \U^x_{\delta} (\o,\nu) := E^{\o}_{x,\nu} \left( (1-\delta) \sum_{n=1}^{\infty} \delta^{n-1} f(\o_n,\x_n)  \right), \quad \delta \in [0,1),
\end{align*}
be the $N$-Ces\`aro and $\delta$-discounted payoffs associated with the initial state $\o$ and behavioral strategy $\nu \in \mathcal{N}$, where $E^{\o}_{x,\nu}$ denotes the expectation operator with respect to $P^{\o}_{x,\nu}$. Note that by the definition of $\G_x$ we have  $\U^x_N (\d_{\l}M, \nu) = \U^x_N (\d^*_{\l}M, \nu)$ and $\U^x_{\delta} (\d_{\l}M, \nu) = \U^x_{\delta} (\d^*_{\l}M, \nu)$ for every $\l \in K$. 

Let $\Upsilon^x_N(\o) := \sup_{\nu \in \mathcal{N}} \U^x_N (\o, \nu)$ and $\Upsilon^x_{\delta}(\o) := \sup_{\nu \in \mathcal{N}} \U^x_{\delta} (\o, \nu)$ denote the $N$-Ces\`aro and $\delta$-discounted value, respectively, of the MDP $\G_x$. By the preceding paragraph we have $\Upsilon^x_N(\d_{\l}M) = \Upsilon^x_N(\d^*_{\l}M)$ and $\Upsilon^x_{\delta}(\d_{\l}M) = \Upsilon^x_{\delta}(\d^*_{\l}M)$ for every $\l \in K$. Hence, we may identify $\Upsilon^x_N$ and $\Upsilon^x_{\delta}$ with their restrictions to $\Delta(K)$ (as opposed to $\Delta(K) \cup \R$).

\begin{assumption}\label{Assump. Exist. Optimal Str.}
For every $\delta \in (0,1)$, $x \in [0,1]$ and $\o \in \Delta(K)$ there exists an optimal strategy, denoted $\hat{\nu}_{\delta}^{x} (\o)$. That is, it holds $\Upsilon^x_{\delta}(\o) = \U^x_{\delta} (\o, \hat{\nu}_{\delta}^{x} (\o))$. 
\end{assumption}

The above assumption is consistent with the discounted values of both Markovian persuasion and zero-sum games
with one-sided information and changing states, having stochastic revelations.

Lastly, to simplify notation, in the case where $x=0$, we will omit the parameter $x$ from all relevant expressions, so that, $P^{\o}_{0,\nu}$, $E^{\o}_{0,\nu}$, $\U^0_N (\o, \nu)$,  $\U^0_{\delta}(\o,\nu)$, $\Upsilon^0_N(\o)$, $\Upsilon^0_{\delta}(\o)$ and $\hat{\nu}_{\delta}^{0} (\o)$, will be denoted by $P^{\o}_{\nu}$, $E^{\o}_{\nu}$, $\U_N (\o, \nu)$,  $\U_{\delta}(\o,\nu)$, $\Upsilon_N(\o)$, $\Upsilon_{\delta}(\o)$, and $\hat{\nu}_{\delta} (\o)$.

\subsubsection{A random-duration valuation for $\G_0$}\label{Subsection Random Duration Valuation}

A new payoff valuation that will play a key role in the subsequent analysis is the random-duration valuation. Such a valuation will be relevant for the case where there are no exogenous revelations, corresponding to the case $x=0$. Formally, let associate with each state $\o$ and behavioral strategy $\nu$ the random-duration payoff, denoted $\U_x^{r\text{-}d}(\o,\nu)$, and defined by
\begin{align}\label{Random Duration Payoff}
    \U_x^{r\text{-}d}(\o,\nu) := E^{\o}_{\nu} \left( \sum_{n=1}^{\infty} f(\o_n,\x_n) \1\{W \geq n\}  \right),
\end{align}
where $W$ is a discrete random variable following a geometric distribution with parameter $x \in (0,1]$, independent of all other instances occurring in $\G_0$.\footnote{To be fully formal, one needs to argue that $W$ can be defined on the space where $E^{\o}_{\nu}$ operates, i.e., infinite histories $(\o_1,\x_1,..., \o_n,\x_n,...)$. We skip this, as we can always enlarge the latter space, in a way that is consistent with $E^{\o}_{\nu}$ and $P^{\o}_{\nu}$.} One can think of the random-duration payoff as describing a scenario in which the play in $\G_0$ terminates at random after $W$ stages, with the decision maker obtaining his sum of the stage payoffs $f(\o_1,\x_1)+...+f(\o_W,\x_W)$ until termination.

Let $\Upsilon^{r\text{-}d}_x (\o) = \sup_{\nu \in \mathcal{N}} \U_x^{r\text{-}d}(\o,\nu)$ be the corresponding random-duration value. As before, since $\Upsilon^{r\text{-}d}_x (\d_{\l}M) =\Upsilon^{r\text{-}d}_x (\d^*_{\l}M)$ for every $\l \in K$, we identify $\Upsilon^{r\text{-}d}_x$ with its restriction to $\Delta(K)$.

In the following lemma we establish a connection between the random duration values $\Upsilon^{r\text{-}d}_x (\o)$ and the $\delta$-discounted values $\upsilon_{\delta} (\o)$ of $\Gamma_0$.

\begin{lemma}\label{Random Duration Value Lemma}
For every $\o \in \Delta(K)$ and $x \in (0,1]$ it holds that  $\Upsilon^{r\text{-}d}_x (\o) = \Upsilon_{1-x} (\o) / x$.
\end{lemma}

\begin{proof}[Proof of \ref{Random Duration Value Lemma}]
    Since $W$ is independent of $(\o_1,\x_1,...,\o_n,\x_n,...)$ we may disintegrate it from \eqref{Random Duration Payoff}, and change the order of summation to obtain,
\begin{equation*}
\begin{split}
\U_x^{r\text{-}d}(\o,\nu) & = \sum_{N=1}^{\infty} (1-x)^{N-1} x \,\, E^{\o}_{\nu} \left(\sum_{n=1}^{N} f(\o_n,\x_n)\right)\\
& = \sum_{n=1}^{\infty} \left(\sum_{N = n}^{\infty} (1-x)^{N-1} x \right)  E^{\o}_{\nu} (f(\o_n,\x_n))\\
& = \frac{1}{x}  \sum_{n=1}^{\infty} x (1-x)^{n-1}\,\, E^{\o}_{\nu} (f(\o_n,\x_n)) =  \U_{1-x}(\o,\nu)/x.
\end{split}
\end{equation*}
Optimizing over $\nu \in \mathcal{N}$ yields the claim.
\end{proof}

A combination of Assumption \ref{Assump. Exist. Optimal Str.} and Lemma \ref{Random Duration Value Lemma} yield the following important strategic corollary. 

\begin{corollary}\label{Corollary - Random Duration Strategy}
For every $\o \in \Delta(K)$ and $x \in (0,1]$ the strategy $\hat{\nu}_{1-x} (\o) \in \mathcal{N}$ is optimal under the random duration payoff $\U_x^{r\text{-}d}$. That is, $\Upsilon^{r\text{-}d}_x (\o) = \U_x^{r\text{-}d}(\o,\hat{\nu}_{1-x} (\o))$. 
\end{corollary}

In words, Corollary \ref{Corollary - Random Duration Strategy} asserts that following an optimal $(1-x)$-discounted strategy in $\G_0$, will be optimal under the random-duration valuation as well, where the duration is determined by a geometric distribution with parameter $x$. 

\subsection{Reducing $\G_x$ to a Markovian persuasion game with stochastic revelations}\label{MDP of MP(x)}

We first consider the game $MP(x)$. Associate with each signaling policy $\s \in \Sigma$ the sequence of priors $(q_n)_{n\geq 1}$, $q_n \in \Delta(K)$ defined by
\begin{equation*}
q_n^{\l} = P^p_{x,\s}\left(X_n=\l\,|\, s_1,r_1,...,s_{n-1},r_{n-1}\right), \quad \forall \l \in K.
\end{equation*}
This sequence describes the prior belief of the receiver in $MP(x)$ regarding $X_n$, prior to obtaining the signal $s_n$ from the sender. 

We now turn to the reduction. First, take $A = S$, so that the actions in $\Gamma_x$ are reduced to $\Delta(S)^K$. The sequence of states $(\o_n)_{n\geq 1}$ in the MDP $\Gamma_x$ will be reduced using the sequence of priors $(q_n)_{n\geq 1}$ as follows: for $n=1$ set $\o_1 = q_1 = p \in \Delta(K)$; for $n\geq 2$, given, $r_{n-1} = \emptyset$, i.e., no revelation has occurred in the $n-1$'st stage of $MP(x)$, set $\o_n = q_n \in \Delta(K)$. If  $r_{n-1} \in K$, set $\o_n = \d^*_{r_{n-1}}M$, as in this case, $X_{n-1}=r_{n-1}$, was revealed to the receiver.

We need to verify that our reduction of $(\o_n)_{n\geq 1}$ satisfies the two items in the transition rule given in \eqref{transition rule}. For this matter let us choose $\xi_n = \s_n (s_1,r_1,...,s_{n-1},r_{n-1},x_n)$. Then, we have that 
\begin{equation}\label{p_n to omega_n}
p_n=\o_n(\xi_n,s) \quad  \text{with probability} \quad \sum_{\l \in K} \o_n^{\l}\cdot \x_n^{\l}(s),
\end{equation} where $p_n$ was defined in \eqref{posterior sequence definition}.\footnote{The careful reader might wonder about the choice $\x_n = \s_n (s_1,r_1...,s_{n-1},r_{n-1},x_n)$. We address it in detail in 
Subsection \ref{Sub.Sec:Equivalence}} Next, in the case of \textbf{no revelation}, i.e., $r_n = \emptyset$, using the fact that given $p_n$, $q_{n+1} = p_n M$, we infer that $\o_{n+1} = p_n M = \o_n(\x_n,s) M$ with probability $(1-x)\sum_{\l \in K} \o_n^{\l}\cdot \x_n^{\l}(s)$, validating the second item in \eqref{transition rule}. As revelations occur in $MP(x)$ after the signaling of the sender, utilizing \eqref{p_n to omega_n} and the definition of $\o_n(\xi_n,s)$, we obtain that given $\o_n$, $q_{n+1}$ will equal $\d_{\l}M$ \textbf{due to a revelation} with probability
\begin{equation*}
\sum_{s \in S} \left( \sum_{i \in K} \o_n^{i}\cdot \x_n^{i}(s) \right)\cdot p_n^{\l} \cdot x  = \sum_{s \in S} \o_n^{\l} \cdot \x_n^{\l}(s) \cdot x = \o_n^{\l} \cdot x.
\end{equation*}
Therefore, $\o_{n+1}$ will equal $\d^*_{\l}M$ in this case with probability $\o_n^{\l} \cdot x$, in consistency with the first item in \eqref{transition rule}.

We complete the reduction by setting $$f(\o,\x) = \sum_{s \in S} \left( \sum_{\l \in K} \o_n^{\l}\cdot \x_n^{\l}(s) \right) u(\o(\x,s)).$$ Put differently, at each stage $n\geq 1$, $f(\o_n,\x_n)$ equals to the expected payoff of $u(p_n)$ given $q_n$.

\subsubsection{Equivalence of values of $MP(x)$ with those of the reduction of $\Gamma_x$ to $MP(x)$}\label{Sub.Sec:Equivalence}

To show that the value $\Upsilon_N^x$ and $\Upsilon_{\d}^x$ under the above-described reduction agree with $v_N^x$ and $v_{\d}^x$ requires additional arguments. First, we argue that the former values are not smaller than the latter. To do so, we describe a machinery that translates each signaling policy $(\s_n)_{n\geq 1}$ to a behavior strategy $(\nu_n)_{n\geq 1}$ in the reduction of $\Gamma_x$. Such translation preserves the payoffs ($N$-Ces\'aro and $\d$-discounted).

The cornerstone argument for such translation is that any history in $\Gamma_x$ of the form $(\o_1,\x_1,...,\o_{n-1},\x_{n-1},\o_n)$ induces a distribution over the space $(K \times S \times \{0,1\})^n$ describing the outcomes $(x_1,s_1,z_1,...,x_n,s_n,z_n)$ in the first $n$ stages of $MP(x)$. To see this, note that the sender, who acts as the decision maker in the MDP $\G_x$, learns of the outcome of the Bernoulli trial by observing whether the MDP $\G_x$ visits the states in $\R$ or not. 

Therefore, given $(\s_n)_{n\geq 1}$, the translation to a behavioral strategy $(\nu_n)_{n\geq 1}$ may be constructed iteratively, such that at any stage $n$, the strategy $\nu_n$ instructs the decision maker in $\Gamma_x$ to first simulate $(x_1,s_1,z_1,...,x_{n-1},s_{n-1}, z_{n-1},x_n)$ from $(\o_1,\x_1,...,\o_{n-1},\x_{n-1},\o_n)$ and then take the action $\x_n = \s_n (x_1,s_1,z_1,...,x_{n-1},s_{n-1}, z_{n-1},x_n)$ (note that $\nu_n (\o_1,\x_1,...,\o_{n-1},\x_{n-1},\o_n)$ is indeed a mixed action over $\Delta(S)^K$).\footnote{We abuse notation, and denote both the random $n$'th signal of the sender and its realization by $s_n$. The appropriate use is made clear based on the context.} 

Conversely, it is not true that each behavior strategy $(\nu_n)_{n\geq 1}$ can be translated to a payoff equivalent signaling policy $(\s_n)_{n\geq 1}$. For instance, if $\nu_1$ mixes between two actions $\xi^*$ and $\xi^{**}$ in $\Delta(S)^K$, the resulting expected payoff for the first stage might not be implemented by a single action $\xi \in \Delta(S)^K$, corresponding to the lottery over signals prescribed by $\s_1$. The key tool that allows us to overcome this issue is the dynamic programming principle for the reduction of $\Gamma_x$. According to such a principle we have for every $\o \in \dk$, $N\geq 1$ and $\delta \in [0,1)$,
\begin{align}
    \Upsilon^x_{N} (\o) & = \sup_{\zeta \in \Delta(\Delta(S)^K)} \int_{\Delta(S)^K} \bigg\lbrace \frac{1}{N} \cdot f(\o, \xi) + \frac{N-1}{N} \cdot x \cdot \sum_{\l \in K} \o^{\l} \cdot \, \Upsilon^x_{N-1}(\delta_{\l}^* M)\nonumber \\
    & \quad + \frac{N-1}{N} \cdot (1-x) \cdot \sum_{s \in S}\, \left[ \sum_{\l \in K} \o^{\l}\cdot \x^{\l}(s) \right] \, \Upsilon^x_{N-1} \left(\o(\x,s) M\right)  \bigg\rbrace d\zeta(\xi),
\end{align}
and 
\begin{align}
    \Upsilon^x_{\d} (\o) & = \sup_{\zeta \in \Delta(\Delta(S)^K)} \int_{\Delta(S)^K} \bigg\lbrace (1-\d) \cdot f(\o, \xi) + \d \cdot x \cdot \sum_{\l \in K} \o^{\l} \cdot \, \Upsilon^x_{\d}(\delta_{\l}^* M)\nonumber \\
    & \quad + \d \cdot (1-x) \cdot \sum_{s \in S}\, \left[ \sum_{\l \in K} \o^{\l}\cdot \x^{\l}(s) \right] \, \Upsilon^x_{\d} \left(\o(\x,s) M\right)  \bigg\rbrace d\zeta(\xi).
\end{align}
However, as both of the above terms under maximization are linear (as a function of $\zeta$), the maximum must be attained at some \textbf{pure} action $\xi \in \Delta(S)^K$. As any pure action may be implemented by a signaling policy $(\s_n)_{n\geq 1}$ we indeed obtain that equivalence of the values $\Upsilon_N^x$ and $\Upsilon_{\d}^x$ with $v_N^x$ and $v_{\d}^x$, under the above reduction of $\Gamma_x$ to $MP(x)$.

\subsection{Reducing $\G_x$ to $RG(x)$}\label{MDP of ZSRG}

The reduction is carried out in steps. First, we reformulate $RG(x)$ in terms of a stochastic game $\hat{\G}_x$, following a similar approach to that in Renault \cite{Renault} (see Remark 5.2 in \cite{Renault}). The \textit{state space} in  $\hat{\G}_x$ is set to equal $\dk \cup \R$. The \textit{action sets} of Players 1 and 2 are $\D(I)^K$ and $\D(J)$, respectively. The \textit{transition rule}, denoted $t_*$, is defined by
\begin{equation*}
    t_*(\o,\x,\b) = t(\o,\x), \quad \forall (\o,\x,\b) \in \{\dk \cup \R\} \times \D(I)^K  \times \D(J), 
\end{equation*}
 In particular, the transition rule is determined by the current state and the action $\x$ of Player 1. Finally, the \textit{payoff} associated with state $\o$ and action pair $(\x,\b) \in \D(I)^K  \times \D(J)$ is given by $G^{\o}[\x,\b]$, where the latter denotes the linear extension of $G^{\l}[i,j]$ to $\{\dk \cup \R\} \times \D(I)^K  \times \D(J)$, formally described by $$G^{\o}[\x,\b]:= \sum_{\l \in K} \o^{\l} \left(\sum_{i \in I}\sum_{j \in J}  \x^{\l}(i)\cdot \b(j) \cdot G^{\l}[i,j]\right).$$

Applying the dynamic programming principle (e.g.\ Theorem IV.3.2, p.\ 184 in \cite{MSZ book}) to the stochastic game $\hat{\G}_x$ we obtain
\begin{align*}
    V_{\delta}^x (\o) & = \max_{\x \in \D(I)^K} \min_{\b \in \D(J)} \Bigg\{ (1-\delta) G^{\o}[\x,\b]\\
    & \quad \quad \quad + \delta \cdot (1-x)\sum_{i\in I} \sum_{\l \in K} \o^{\l}\cdot \x^{\l}(i) \cdot V_{\delta}^x\big(\o(\x,i)M\big) + \delta x \sum_{\l \in K} \o^{\l} V_{\delta}^x (\d^*_{\l}M) \Bigg\}.
\end{align*}
The above formula implies that for any discount factor $\delta$, the best response by Player 2 to an action $\x \in \D(I)^K$ of Player 1 is an element of $ \argmin_{\b \in \D(J)} G^{\o}[\x,\b]$.

The reduction of $\hat{\G}_x$ to $\G_x$ is now achieved by taking $A = I$, and $f$ to be given by $f (\o,\x) = G^{\o}[\x, \b_{\x}]$, where for each $\x \in \D(A)^K$, $\b_{\x}$ is an arbitrary choice of element in $ \argmin_{\b \in \D(J)} G^{\o}[\x,\b]$.

The value equivalence under such a reduction follows by a parallel use of the dynamic programming principle, to the one described in Subsection \ref{Sub.Sec:Equivalence}.

\section{Proofs of Theorem \ref{Thm1} and \ref{Thm:RG-1}}\label{Proof of Uniform and Asym Values}

\subsection{The plan of the proof}\label{Subsec MDP plan}

The proofs of Theorem \ref{Thm1} and \ref{Thm:RG-1} will follow from an analogous general result for the MDP $\G_x$. Namely, in this section, we will prove the following Theorem.

\begin{theorem}\label{Uniform Value for MDP}
The $\delta$-discounted values $\Upsilon^x_{\delta} (\cdot)$ of $\G_x$ converge uniformly on $\Delta(K)$ as $\delta \to 1^-$ to
\begin{align*}
\V^*(\o,x) =  \sum_{i=1}^r \left[ \o(C_i) + \sum_{j \in T} \o^j \cdot M[j \to C_i] \right] \left( \sum_{\l \in C_i} \pi_{C_i}^{\l} \cdot \Upsilon_{1-x}(\emph{\textbf{m}}_{\l})   \right).
\end{align*}
Moreover, there exists a behavioral strategy $\nu^* \in \mathcal{N}$ such that for every $\o \in \Delta(K)$ it holds $\lim_{\delta \to 1^-}  \U^x_{\delta} (\o, \nu^*) = \V^*(\o,x)$.
\end{theorem}

The proof of Theorem \ref{Uniform Value for MDP} will be achieved by describing a behavioral strategy $\nu^* \in \mathcal{N}$, independent of the initial state $\o_1 = \o \in \Delta(K)$, satisfying the following two identities:
\begin{equation}\label{Eq. Uniform1}
\U^x_N (\o, \nu) \leq \U^x_N (\o, \nu^*) + o_N(1), \quad  \forall \nu \in \mathcal{N}, \,\,\, \forall \o \in \Delta(K),
\end{equation}
and 
\begin{equation}\label{Eq. Uniform2}
\vert \U^x_N (\o, \nu^*) - \V^*(\o,x) \vert \leq o_N(1), \quad \forall \o \in \Delta(K),
\end{equation}
where by $o_N(1)$ we refer to a sequence of non-negative numbers $(c_N)_{N\geq 1}$ converging to $0$ as $N \to \infty$.\footnote{The subscript $N$ is added to the typical $o(1)$ notation, as the proof will include sequences parametrized by various indices. Therefore, the notation $o_N(1)$ clarifies that the sequence vanishes as its denoted index, $N$, grows arbitrarily large.} An element $c_N$ of an $o(1)$ sequence might depend on $x$ which will be fixed throughout the proof; however, $c_N$ cannot depend on a given state $\o \in \Delta(K)\cup \R$, action $\x \in \Delta(A)^K$ nor a behavioral strategy $\nu \in \mathcal{N}$.

The fact that Eqs.\ (\ref{Eq. Uniform1}) and (\ref{Eq. Uniform2}) are sufficient for deducing Theorem \ref{Uniform Value for MDP} is due to two separate Tauberian type theorems. The first, titled Uniform Tauberian Theorem for Dynamic Programming, was proved by Lehrer and Sorin in \cite{Lehrer}, and implies that for general Markov decision processes, if either one of the $\delta$-discounted or $N$-Ces\`aro values admits a uniform limit, this limit must be a uniform limit for both valuations. The second, often titled Abelian Theorem (e.g., Theorem 2.1 in \cite{SolanB}), ensures that if $(z_n)_{n\geq 1}$ is a bounded sequence of non-negative numbers whose $N$-Ces\`aro averages converge to $z_* \geq 0$ (as $N \to \infty$), then the $\delta$-discounted averages of $(z_n)_{n\geq 1}$ converge (as $\d \to 1^-)$ to $z_*$ as well.

The Uniform Tauberian Theorem for Dynamic Programming is employed in our context as follows. As  $\U^x_N (\cdot, \nu^*) \leq \Upsilon^x_N(\cdot)$, Eq.\ (\ref{Eq. Uniform1}) implies that $\vert \Upsilon^x_N (\cdot) - \U^x_N (\cdot, \nu^*)\vert \leq o(1)$. This in turn with Eq.\ (\ref{Eq. Uniform2}) yields that $\vert \Upsilon^x_N (\cdot) - \V^*(\cdot,x)\vert \leq o(1)$, so that $\Upsilon^x_N(\cdot)$ converges uniformly on $\Delta(K)$ to $\V^*(\cdot,x)$, and thus so does $\Upsilon^x_{\delta} (\cdot)$. 

As for the relevance of the Abelian Theorem, note that for every $\o \in \Delta(K)$,  $\U^x_N (\o, \nu^*)$ is the $N$-Ces\`aro average of the sequence $(z_n (\o))_{n\geq 1}$, where $z_n (\o) := E^{\o}_{x,\nu^*} f(\o_n,\x_n)$, whereas $\U^x_{\delta} (\o, \nu^*)$ is the $\delta$-discounted average of $(z_n (\o))_{n\geq 1}$. Therefore, as by Eq.\ (\ref{Eq. Uniform2}) $\U^x_N (\o, \nu^*)$ converges to $\V^*(\o,x)$ for every $\o \in \dk$, by the Abelian Theorem $\U^x_{\delta} (\o, \nu^*)$ converges to $\V^*(\o,x)$ for every $\o \in \dk$ as well. Thus, altogether we have established the sufficiency of  Eqs.\ (\ref{Eq. Uniform1}) and (\ref{Eq. Uniform2}) for the deduction of Theorem \ref{Uniform Value for MDP}.

The first step toward the proofs of  Eqs.\ \eqref{Eq. Uniform1} and \eqref{Eq. Uniform2} will require us to introduce a probabilistic setup for $\G_x$, which will allow us to incorporate relevant tools from the field of Probability Theory.

\subsection{Probabilistic approach to $\G_x$}\label{Subsec Prob Approach}

Our starting point is to introduce a general form of payoff, which takes into account a \textit{random time} $T$. Such a random time, which is assumed to take positive integer values, is defined on the space consisting of infinite sequences $(\o_1,\x_1,...,\o_n,\x_n,...)$ equipped with the probability measure $P^{\o}_{x,\nu}$. Furthermore, we assume that for each $\nu \in \mathcal{N}$, the distribution of $T$ under $P^{\o}_{x,\nu}$ is the same, i.e., $T$ is independent of the strategy $\nu$. Define for every $\o \in \dk$ and $\nu \in \mathcal{N}$
\begin{align*}
    \W_T (\o,\nu) = E^{\o}_{x,\nu} \left( \sum_{n=1}^T f(\o_n,\x_n) \right)
\end{align*}
to be the expected sum of stage payoffs under $\nu$ along the first $T$ stages in $\G_x$. In particular, if $T$ is constant, i.e., $T=N$ for some $N \in \N$, then $\W_T (\o,\nu) = N\cdot \U^x_N (\o, \nu)$. It follows from the definition that for any two random times $T,T'$  it holds 
\begin{align}\label{Eq. T dist 1}
    \vert \W_T (\o,\nu)  - \W_{T'} (\o,\nu) \vert  \leq \Vert f \Vert E^{\o}_{x,\nu} \, \vert T  - T' \vert, \quad \forall \o \in \dk,\, \forall \nu \in \mathcal{N}.
\end{align}

We move on by defining the following sequence of random stopping times:
\begin{equation*}
\begin{split}
\k_1 & := \inf\,\{\,n > 1\,:\, \o_n \in \R \},  \\	
\k_n & := \inf\,\{\, n \geq 1 \,:\, \o_{\k_1+...+\k_{n-1}+n} \in \R \, \}, 	\	\	 n \geq 2. \nonumber
\end{split}
\end{equation*}
By definition, $\k_1$ describes the first stage at which $\G_x$ visits the set of states $\R$. Note that since by definition the initial state $\o_1$ is assumed to be in $\Delta(K)$, $\k_1$ must be greater than $1$. Thereafter, for each $n\geq 2$, $\k_n$ counts the number of stages between successive visits of $\G_x$ to the set of states $\R$.

The distribution of the random variables $(\k_n)_{n\geq 1}$, defined on the probability space consisting of infinite sequences $(\o_1,\x_1,...,\o_n,\x_n,...)$ equipped with the probability measure $P^{\o}_{x,\nu}$, where $\nu \in \mathcal{N}$, is independent of $\nu$. In fact, we have that $\k_1-1, \k_2,...,\k_n,...$ are i.i.d.\ geometric random variables with parameter $x$. It is for this reason that we do not specify $\nu$ when referring to the visit times $(\k_n)_{n\geq 1}$. 

Consider now for each positive integer $m$ the random time 
\begin{align*}
    T_m = \k_1 + ...+ \k_m-1,
\end{align*}
describing the number of stages before the $m$'th visit of $\G_x$ in $\R$. Since $\k_1-1, \k_2,...,\k_n,...$ are i.i.d.\ geometric random variables with parameter $x$, we have for any $\nu \in \mathcal{N}$ that
\begin{align}\label{Id T_m mean}
    E^{\o}_{x,\nu}\, T_m =  \frac{m}{x}, 
\end{align}
and 
\begin{align}\label{Id T_m var}
    E^{\o}_{x,\nu}\, (T_m - E^{\o}_{x,\nu}\, T_m)^2 = \frac{m(1-x)}{x^2},
\end{align}
where we used the fact that a geometric distribution with parameter $x$ has mean $1/x$ and variance $(1-x)/x^2$, and the fact that the variance of a sum of independent random variables equals the respective sum of variances.

Consider the sequence of positive integers $(L_m)_{m\geq 1}$ defined by $L_m = \lfloor m/x\rfloor$. A key property of $(L_m)_{m\geq 1}$ that will play an important role in our proof is that it has \textit{bounded increments}, i.e., 
\begin{align}\label{Eq. bounded increments}
    L_{m+1} - L_m \leq  \frac{m+1}{x} - \frac{m}{x}  =  \frac{1}{x}, \quad \forall m \geq 1.
\end{align}

The following lemma establishes the asymptotic proximity between the payoffs $\W_{T_m}/L_m$ and $\U^x_{L_m}$.

\begin{lemma}\label{Lemma Approx. 1}
    For every $\o \in \dk$ and $\nu \in \mathcal{N}$ it holds that 
\begin{align*}
    \displaystyle  \left\vert \frac{1}{L_m}\W_{T_m} (\o,\nu)  - \frac{1}{L_m}\W_{L_m} (\o,\nu) \right\vert \leq o_m (1).
\end{align*}
\end{lemma}

\begin{proof}[Proof of Lemma \ref{Lemma Approx. 1}]
   Using the inequality in \eqref{Eq. T dist 1} we obtain
    \begin{align}\label{Eq. T dist 2}
       \displaystyle  \left\vert \frac{1}{L_m}\W_{T_m} (\o,\nu)  - \frac{1}{L_m} \W_{L_m}(\o, \nu) \right\vert  & \leq \frac{1}{L_m} \Vert f \Vert \cdot  E^{\o}_{x,\nu}\, \vert T_m - L_m \vert \\
       & = \Vert f \Vert \cdot  E^{\o}_{x,\nu}\, \left\vert \frac{T_m}{\lfloor E^{\o}_{x,\nu}\, T_m \rfloor } - 1 \right\vert, \nonumber
    \end{align}
    where the last equality follows from the fact that by the definition of $L_m$ and identity \eqref{Id T_m mean} it holds $L_m = \lfloor E^{\o}_{x,\nu}\, T_m \rfloor$. Next, as 
    \begin{align*}
        \left\vert \frac{T_m}{\lfloor E^{\o}_{x,\nu}\, T_m \rfloor } - 1 \right\vert & \leq \left\vert \frac{T_m}{E^{\o}_{x,\nu}\, T_m} - 1 \right\vert + \left( \frac{1}{\lfloor E^{\o}_{x,\nu}\, T_m \rfloor}  - \frac{1}{ E^{\o}_{x,\nu}\, T_m } \right) T_m \\ 
        & \leq \left\vert \frac{T_m}{E^{\o}_{x,\nu}\, T_m} - 1 \right\vert + \frac{1}{ (\lfloor E^{\o}_{x,\nu}\, T_m \rfloor)^2} \cdot T_m,
    \end{align*}
we may further relax the inequality in Eq.\ \eqref{Eq. T dist 2} to obtain 
\begin{align}\label{Eq. T dist 3}
    \left\vert \frac{1}{L_m}\W_{T_m} (\o,\nu)  - \frac{1}{L_m}\W_{L_m} (\o,\nu) \right\vert & \leq \Vert f \Vert \cdot  E^{\o}_{x,\nu}\, \left\vert \frac{T_m}{E^{\o}_{x,\nu}\, T_m} - 1 \right\vert \nonumber \\
    & \quad  + \Vert f \Vert \frac{E^{\o}_{x,\nu}\, T_m}{(\lfloor E^{\o}_{x,\nu}\, T_m \rfloor)^2}.
\end{align}
Applying the Cauchy-Schwarz inequality and identity \eqref{Id T_m var} we have that 
\begin{align*}
    E^{\o}_{x,\nu}\, \left\vert \frac{T_m}{E^{\o}_{x,\nu}\, T_m} - 1 \right\vert & \leq  \sqrt{E^{\o}_{x,\nu}\, \left(\frac{T_m}{E^{\o}_{x,\nu}\, T_m} - 1 \right)^2}\\
    & = \frac{1}{E^{\o}_{x,\nu}\, T_m} \sqrt{E^{\o}_{x,\nu}\, (T_m - E^{\o}_{x,\nu}\, T_m)^2}\\
    & = \frac{1}{m/x} \sqrt{\frac{m(1-x)}{x^2}} = \sqrt{\frac{1-x}{m}}.
\end{align*}
Combining the above with Eq.\ \eqref{Eq. T dist 3} and the fact that $\lfloor E^{\o}_{x,\nu}\, T_m \rfloor \geq m/x -1$, we obtain 
 \begin{align*}
     \left\vert \frac{1}{L_m}\W_{T_m} (\o,\nu)  - \frac{1}{L_m}\W_{L_m} (\o,\nu) \right\vert \leq \Vert f \Vert \left( \sqrt{\frac{1-x}{m}} +  \frac{m/x}{(m/x-1)^2} \right)
 \end{align*}
which is sufficient for deducing the statement of the lemma. 
\end{proof}

Let us next define for each $m\geq 1$
\begin{align*}
    \mathcal{Q}_m (\o, \nu) := \W_{T_m} (\o,\nu) - \W_{T_1} (\o,\nu), \quad \forall\o \in \dk, \, \forall \nu \in \mathcal{N}.
\end{align*}

In different terms, $\mathcal{Q}_m (\o, \nu)$ describes the expected sum of stage payoffs beginning from the stage where the first `revelation' (visit in $\R$) takes place (i.e., $\k_1 = T_1 + 1$) and ending at the stage preceding the $m$'th `revelation' ($\k_1+...+\k_m = T_m+1$).

The following approximation will be used extensively.

\begin{claim}\label{Claim Approx. 1}
     For every $\o \in \dk$ and $\nu \in \mathcal{N}$ it holds that 
\begin{align*}
    \displaystyle  \left\vert \frac{1}{L_m}\mathcal{Q}_m (\o, \nu)  - \U^x_{L_m}(\o, \nu) \right\vert \leq o_m (1).
\end{align*}
\end{claim}

\begin{proof}[Proof of Claim \ref{Claim Approx. 1}]
    As by definition
    \begin{align*}
       \displaystyle  \left\vert \mathcal{Q}_m(\o, \nu) -\W_{T_m} (\o, \nu)  \right\vert  & = \left\vert E^{\o}_{x,\nu} \left( \sum_{n=1}^{\k_1-1} f(\o_n,\x_n) \right) \right\vert \leq \Vert f \Vert E^{\o}_{x,\nu} (\k_1-1) = \frac{\Vert f \Vert}{x},
    \end{align*}
    we obtain that $ \left\vert \mathcal{Q}_m(\o, \nu)/L_m -\W_{T_m} (\o, \nu)/L_m  \right\vert  \leq o_m(1)$, which in conjunction with Lemma \ref{Lemma Approx. 1} and the fact that $\W_{L_m} (\o, \nu)/L_m =  \U^x_{L_m}(\o, \nu) $  yields the desired result.
\end{proof}

Upon completing the required approximations for the subsequent analysis, we are in a position to introduce the strategy $\nu^*$, and thereafter prove the two required identities, given in relations \eqref{Eq. Uniform1} and \eqref{Eq. Uniform2}, with $\nu^*$.

\subsection{The strategy $\nu^*$}\label{Subsec nu-star}

We begin with the formal definition of $\nu^* \in \mathcal{N}$ and succeed with its motivation.

The strategy $\nu^*$ is defined by stages as follows:
\begin{itemize}
    \item Play arbitrary until stage $T_1 + 1$.
    \item For every $1 \leq m$, starting from stage $T_m+1$, conditional on $\o_{T_m + 1} = \delta_{\l}^*M$, follow the strategy $\hat{\nu}_{1-x}(\delta_{\l}M)$, until stage $T_{m+1}$ (included).
\end{itemize}

Two immediate remarks are the following. First, $\nu^*$ is independent of $\o_1$. This is due to the definition of $\nu^{*}$ and the fact that $T_1+1=\k_1 > 1$. Secondly, note that the decision maker in $\G_x$ is not aware in real time that the current stage is of the form $T_{m+1}$, as he is unable to know whether tomorrow, i.e., at stage $T_{m+1}+1 = \k_1 + \cdots + \k_m$, a revelation will indeed occur. 

However, the strategy $\nu^*$ is still well defined, as one may interpret the decision as to whether to follow the strategy $\hat{\nu}_{1-x}(\delta_{\l}M)$ as one that is taken at the start of each stage following stage $T_m+1$. Therefore, under such interpretation, only at the start of stage $T_{m+1}+1$, which is known to the decision maker, the decision is to stop following $\hat{\nu}_{1-x}(\delta_{\l}M)$, and to start following $\hat{\nu}_{1-x}(\delta_{i}M)$, given that $\o_{T_{m+1}+1} = \delta_{i}^*M$.

\subsubsection{The strategic motivation behind $\nu^*$}

The strategy $\nu^*$ seeks to capitalize on the renewal of information each time an exogenous revelation takes place in $\G_x$. Indeed, upon an exogenous revelation, all previous learning in $\G_x$, does not affect future learning. Therefore, $\nu^*$ is instructed to play independently between any two successive exogenous revelations.

The duration between two successive revelations, described by the stages $T_m + 1,..., T_{m+1}$, $m\geq 1$, admits a geometric distribution with parameter $x$. Moreover, knowing that $\o_{T_m + 1} = \delta_{\l}^*M$, and that no additional exogenous revelations will occur until stage $T_{m+1}+1$, in essence, along $T_m + 1,..., T_{m+1}$ the decision maker faces $\G_0$ with a random duration valuation, starting from the prior $\delta_{\l}M$.

The study of $\G_0$ with a random duration valuation, as detailed in Subsection \ref{Subsection Random Duration Valuation}, suggests that starting from $\o_{T_m + 1} = \delta_{\l}^*M$, it is optimal to follow $\hat{\nu}_{1-x}(\delta_{\l}M)$ (see Corollary \ref{Corollary - Random Duration Strategy}). Thus, in summary, $\nu^*$ acts as if it faces an \textbf{independent sequence of random duration games with no exogenous revelations}, and wishes to maximize its expected sum of stage payoffs in each one of those games.  

We are in a position to deduce the sufficient conditions for Theorem \ref{Uniform Value for MDP} given in Eqs.\ \eqref{Eq. Uniform1} and \eqref{Eq. Uniform2}. We begin with the former.

\subsection{Proof of Eq.\ \eqref{Eq. Uniform1}}\label{Subsec nu-star optimality}

As preparation for the proof, we associate with each $\nu \in \mathcal{N}$, $\o \in \dk$ and $r\geq 1$ a behavioral strategy $\nu[\o,r] \in \mathcal{N}$. The strategy $\nu[\o,r]$ chooses at each stage $n$, given the current state $\o_n$, a mixed action $\zeta_n$ over $\Delta(A)^K$, distributed according to the conditional distribution under $P^{\o}_{x,\nu}$ of $\x_{n+r-1}$ given $\o_n$, where we recall $\x_{n+r-1}$ denotes the action at the $(n+r-1)$'st stage of $\G_x$. 

The introduction of $\nu[\o,r] \in \mathcal{N}$ is useful, as it allows us to rewrite for each $1 \leq i \leq m -1$ the expected sum of payoffs along stages $T_i, ..., T_{i+1}-1$ in terms of the random duration payoff $\U_x^{r\text{-}d}$. Formally, we have each $1 \leq i \leq m -1$,
\begin{align*}
    E_{x,\nu}^{\o} \left( \sum_{n = T_i + 1}^{ T_{i+1}} f(\o_n,\x_n)\,|\, T_i+1 = r, \o_{T_i+1} = \delta_{\l}^*M \right) = \U_x^{r\text{-}d}(\delta_{\l}M, \nu[\o,r]) 
\end{align*}
A combination of the above, with the definition of $\nu^*$, and Corollary \ref{Corollary - Random Duration Strategy} yields 
\begin{align}\label{Eq. Optimality of nu* 1}
    E_{x,\nu}^{\o} \left( \sum_{n = T_i + 1}^{ T_{i+1}} f(\o_n,\x_n)\,|\, T_i+1 , \o_{T_i+1}  \right) \leq  E_{x,\nu^*}^{\o} \left( \sum_{n = T_i + 1}^{ T_{i+1}} f(\o_n,\x_n)\,|\, T_i+1 , \o_{T_i+1}  \right)
\end{align}
for every $1\leq i \leq m -1$ and every $\nu \in \mathcal{N}$. Next, we note that for every $i\geq 1$ the marginal distribution of $(T_i+1 , \o_{T_i+1})$ is independent of $\nu \in \mathcal{N}$. Indeed, the distribution of $T_i+1$, which is determined by the distributions of $\k_1,...,\k_i$ is independent of $\nu$. Thereafter, by the definition of $\G_x$, we have for every $\l \in K$ and $\nu \in \mathcal{N}$ that
\begin{align}\label{Eq. marginal state expectation}
& P_{x,\nu}^{\o} \left(\o_{T_i+1} = \delta_{\l}^* M\, |\, T_i+1 = r \right) =  E_{x,\nu}^{\o} \left( \1\{\o_{T_i+1} = \delta_{\l}^* M \} \,|\, T_i+1 = r \right) \nonumber \\
   & \quad \quad \quad \quad \quad \quad =  E_{x,\nu}^{\o} \left( E_{x,\nu}^{\o} \left( \1\{\o_{T_i+1} = \delta_{\l}^* M \} \,|\, T_i+1 = r,\, \o_{r-1} \right) \right)\\
   & \quad \quad \quad \quad \quad \quad = E_{x,\nu}^{\o} \left(\o_{r-1}^{\l}\right) = (\o M^{r-2})^{\l} \nonumber
\end{align}
where the last equality follows from the properties of the transition rule of $\G_x$. Therefore, we may use the law of total expectation on both sides in \eqref{Eq. Optimality of nu* 1} and sum over $i=1,...,m-1$ to obtain
\begin{align*}
    \mathcal{Q}_m(\o,\nu) \leq \mathcal{Q}_m(\o,\nu^*), \quad \forall \nu \in \mathcal{N}.
\end{align*}
This, in turn with Claim \ref{Claim Approx. 1} implies that 
\begin{align}\label{Eq. Almost Suff 1}
 \U_{L_m}^x (\o,\nu) \leq \U_{L_m}^x (\o,\nu^*) + o_m(1), \quad \forall \o \in \dk, \, \forall \nu \in \mathcal{N}.   
\end{align}
Let us now consider a general $N \in \N$ satisfying $N> L_1$. Define $m(N)$ to be the unique positive integer for which $L_{m(N)} < N \leq L_{m(N)+1}$. Since $$N\, \U_{N}^x (\o,\nu) = L_{m(N)}\, \U_{L_{m(N)}}^x (\o,\nu) + \sum_{n= L_{m(N)}+1}^N E_{x,\nu}^{\o}(f(\o_n,\x_n))$$ for every $\o \in \dk$ and $\nu \in \mathcal{N}$, one has 
\begin{align}\label{Eq. bounded increments app}
\left\vert \U_{N}^x (\o,\nu) - \U_{L_{m(N)}}^x (\o,\nu) \right\vert & =  \Bigg\vert \frac{1}{N}\sum_{n= L_{m(N)}+1}^N E_{x,\nu}^{\o}(f(\o_n,\x_n)) \nonumber\\
& \quad \quad \quad - \left(\frac{N-L_{m(N)}}{N}\right) \, \U_{L_{m(N)}}^x (\o,\nu) \Bigg\vert  \\
& \leq \Vert f \Vert \cdot \frac{N-L_{m(N)}}{N} + \Vert f \Vert \cdot \frac{N-L_{m(N)}}{N} \nonumber\\
& \leq 2 \Vert f \Vert \cdot \frac{L_{m(N)+1}- L_{m(N)}}{L_{m(N)}} \leq  2 \Vert f \Vert \cdot \frac{1/x}{L_{m(N)}}, \nonumber
\end{align}
where the last inequality follows from the bounded increments property of $(L_m)_{m\geq 1}$ given in Eq.\ \eqref{Eq. bounded increments}. Therefore, we obtain that $\left\vert \U_{N}^x (\o,\nu) - \U_{L_{m(N)}}^x (\o,\nu) \right\vert \leq o_N(1)$ for every $\o \in \dk$ and $\nu \in \mathcal{N}$, which together with the inequality in \eqref{Eq. Almost Suff 1} implies that
\begin{align*}
     \U_{N}^x (\o,\nu) \leq \U_{N}^x (\o,\nu^*) + o_N(1), \quad \forall \o \in \dk, \, \forall \nu \in \mathcal{N},
\end{align*}
as desired.

\subsection{Proof of Eq.\ \eqref{Eq. Uniform2}}\label{Subsec nu-star payoff}

As we've seen in the proof of Eq.\ \eqref{Eq. Uniform1}, the bounded increments property of $L_m$ implies that to prove Eq.\ \eqref{Eq. Uniform2} it is sufficient to show that 
\begin{align*}
    \left\vert \,\U_{L_{m}}(\o,\nu^*) - \V^* (\o,x) \right\vert \leq o_m(1), \quad  \forall \o \in \dk.
\end{align*}
Furthermore, the approximation in Claim \ref{Claim Approx. 1} reduces the problem to showing that
\begin{align}\label{Eq. nstar reduction}
   \left\vert \frac{1}{L_m}\mathcal{Q}_m(\o,\nu^*) - \V^*(\o,x) \right\vert \leq o_m(1), \quad \forall \o \in \dk.
\end{align}
By the definition of $\nu^*$, Lemma \ref{Random Duration Value Lemma}, and Corollary \ref{Corollary - Random Duration Strategy} we have for any $1 \leq i \leq m-1$:
\begin{align*}
     E_{x,\nu^*}^{\o} \left( \sum_{n = T_i + 1}^{ T_{i+1}} f(\o_n,\x_n)\,|\, T_i+1 , \o_{T_i+1}  \right) = \Upsilon_{1-x} (\o_{T_i+1})/x.
\end{align*}
Taking expectations, summing over $i = 1,...,m-1$, and dividing by $L_m$ we obtain
\begin{align}\label{Eq. nstar1}
   \frac{1}{L_m} \mathcal{Q}_m(\o,\nu^*) & = E_{x,\nu^*}^{\o} \left( \frac{1}{L_m} \sum_{i=1}^{m-1} \Upsilon_{1-x} (\o_{T_i+1})/x  \right)\nonumber \\
   & = E_{x,\nu^*}^{\o} \left( \frac{1}{L_m} \sum_{i=1}^{m-1} \sum_{\l \in K} \1\{ \o_{T_i+1} = \delta_{\l}^*M \}  \Upsilon_{1-x} (\delta_{\l}M) /x  \right) \\
   & = \sum_{\l \in K}  E_{x,\nu^*}^{\o} \left(\frac{1}{L_m} \sum_{i=1}^{m-1} \1\{ \o_{T_i+1} = \delta_{\l}^*M \} \right)  \Upsilon_{1-x} (\delta_{\l}M) /x \nonumber 
\end{align}
We now observe that for each $\l \in K$ we have 
\begin{align}\label{Eq. nstar2}
    E_{x,\nu^*}^{\o} \left(\frac{1}{L_m} \sum_{i=1}^{m-1} \1\{ \o_{T_i+1} = \delta_{\l}^*M \} \right) & = E_{x,\nu^*}^{\o} \left(\frac{1}{L_m} \sum_{n=1}^{T_m} \1\{ \o_{n} = \delta_{\l}^*M \} \right) \nonumber \\ 
    & = \frac{1}{L_m} \W^{g_{\l}}_{T_m} (\o, \nu^*),
\end{align}
where $\W^{g_{\l}}_{T_m} (\o, \nu^*)$ is a private choice of $\W_{T_m}(\o, \nu^*)$ for the case where $f = {g_{\l}}$, and ${g_{\l}}$ is defined by ${g_{\l}}(\o,\x) := \1\{\o = \delta_{\l}^*M\}$ for all $(\o,\x)$. To elaborate, the expression $\W_{T_m} (\o, \nu^*)$ takes the payoff $f$ of the MDP $\G_x$ as a primitive in an implicit form. In $\W^{g_{\l}}_{T_m} (\o, \nu^*)$ we specify explicitly the value of the payoff $f$ we are interested in. Thus, combining Eqs.\ \eqref{Eq. nstar1} with \eqref{Eq. nstar2} we obtain the following identity:
\begin{align}\label{Eq. nstar3}
    \frac{1}{L_m} \mathcal{Q}_m(\o,\nu^*)= \sum_{\l \in K} \frac{\W^{g_{\l}}_{T_m} (\o, \nu^*)}{L_m} \cdot \Upsilon_{1-x} (\delta_{\l}M) /x.
\end{align}

Next, as the analysis in Section \ref{Subsec Prob Approach} was performed for general (bounded) $f$, we may utilize Lemma \ref{Lemma Approx. 1} for $\W^{g_{\l}}_{T_m} (\o, \nu^*)$ to obtain that for each $\l \in K$ it holds
\begin{align*}
    \left\vert \frac{\W^{g_{\l}}_{T_m} (\o, \nu^*)}{L_m} - \frac{\W^{g_{\l}}_{L_m} (\o, \nu^*)}{L_m} \right\vert \leq o_m(1),
\end{align*}
which in turn with Eq.\ \eqref{Eq. nstar3} and the fact that $\Upsilon_{1-x} (\delta_{\l}M) /x \leq \Vert f \Vert /x $ for every $\l$ implies that
\begin{align}\label{Eq. General Mono Approx.}
    \left\vert  \frac{1}{L_m} \mathcal{Q}_m (\o, \nu^*) -  \sum_{\l \in K} \frac{\W^{g_{\l}}_{L_m} (\o, \nu^*)}{L_m} \cdot \Upsilon_{1-x} (\delta_{\l}M) /x    \right\vert \leq o_m(1).
\end{align}
In view of the latter relation and the definition of $\V^*(\o,x)$, in order to establish the sufficient condition for the proof of Eq.\ \eqref{Eq. Uniform2} given in \eqref{Eq. nstar reduction}, we will show that for every $i=1,...,r$ and $\l \in C_i$ it holds that
\begin{align}\label{Eq. nstar sufficient}
    \left\vert \frac{\W^{g_{\l}}_{L_m} (\o, \nu^*)}{L_m} - \left( \o(C_i) + \sum_{j \in T} \o^j \cdot M[j \to C_i] \right) \cdot \pi_{C_i}^{\l} \cdot x \right\vert \leq o_m(1),
\end{align}
whereas for every $\l \in T$ it holds that
\begin{align}\label{Eq. nstar sufficient 2}
    \left\vert \frac{\W^{g_{\l}}_{L_m} (\o, \nu^*)}{L_m} \right\vert \leq o_m(1).
\end{align}
To establish both of the statements, we observe that for every $\l \in K$ we have by definition of $\W^{g_{\l}}_{L_m}(\o, \nu^*)$ that
\begin{align}\label{Eq. nstar4}
    \frac{\W^{g_{\l}}_{L_m} (\o, \nu^*)}{L_m}  & = E_{x,\nu^*}^{\o} \left( \frac{1}{L_m}\sum_{n=1}^{L_m}  \1\{\o_n = \delta_{\l}^* M \} \right) \nonumber \\
    & = \frac{1}{L_m} \sum_{n=2}^{L_m} P_{x,\nu^*}^{\o}(\{ \o_n = \delta_{\l}^* M \}) \nonumber \\
    & = \frac{1}{L_m} \sum_{n=2}^{L_m} x \cdot (\o M^{n-2})^{\l} \\
    & = x \cdot \sum_{j \in K} \o^j \cdot \left(  \frac{1}{L_m} \sum_{n=2}^{L_m} (\delta_{\{j\}}M^{n-2})^{\l} \right), \nonumber 
\end{align}
where the previous to last equality follows by arguments parallel to those presented in Eq.\ \eqref{Eq. marginal state expectation}. Consider now the case where $\l \in T$. As $\l$ is transient, i.e., cannot be visited infinitely many times, we have that for every $j \in K$ it holds 
\begin{align}\label{Eq. nstar5}
    \left\vert \frac{1}{L_m} \sum_{n=2}^{L_m} (\delta_{\{j\}}M^{n-2})^{\l} \right\vert = o_m(1).
\end{align}
Therefore, as $\o^j \leq 1$ for every $j \in K$, Eqs.\ \eqref{Eq. nstar4} and \eqref{Eq. nstar5} imply that $\left\vert \W^{g_{\l}}_{L_m} (\o, \nu^*)/L_m \right\vert \leq o_m(1)$, recovering the statement in \eqref{Eq. nstar sufficient 2}.  Next, to show the statement in \eqref{Eq. nstar sufficient}, let us fix $i \in 1,...,r$ and $\l \in C_i$. The Ergodic Theorem for irreducible Markov Chains (e.g., Theorem C.1 in \cite{Peres}) implies that 
\begin{align}\label{Eq. nstar6}
    \left\vert \frac{1}{L_m} \sum_{n=2}^{L_m} (\delta_{\{j\}}M^{n-2})^{\l} - \pi_{C_i}^{\l} \right\vert = o_m(1), \quad \forall j \in C_i.
\end{align}
Moreover, this theorem also implies that
\begin{align}\label{Eq. nstar7}
    \left\vert \frac{1}{L_m} \sum_{n=2}^{L_m} (\delta_{\{j\}}M^{n-2})^{\l} - M[j \to C_i] \cdot \pi_{C_i}^{\l} \right\vert = o_m(1), \quad \forall j \in T.
\end{align}
as the Markov chain starting from $j \in T$, will be absorbed in finite time in $C_i$ with probability $M[j \to C_i]$, after which the frequency of visits to state $\l$ will tend to $\pi_{C_i}^{\l}$. Lastly, for $j \notin C_i \cup T$, there is a zero probability to visit state $\l$ starting from state $j$, and thus it holds that 
\begin{align}\label{Eq. nstar8}
   \frac{1}{L_m} \sum_{n=2}^{L_m} (\delta_{\{j\}}M^{n-2})^{\l} = 0, \quad \forall m\geq 1, \,\,\, \forall j \notin C_i \cup T.
\end{align}
As $\o^j \leq 1$ for every $j \in K$, we infer with the help of Eqs.\ \eqref{Eq. nstar4}, \eqref{Eq. nstar6}, \eqref{Eq. nstar7}, and \eqref{Eq. nstar8} that 
\begin{align*}
  o_m(1) & \geq \left\vert  \frac{\W^{g_{\l}}_{L_m} (\o, \nu^*)}{L_m} -  \sum_{j \in C_i} \o^j \cdot \pi_{C_i}^{\l} \cdot x - \sum_{j \in T} \o^j \cdot  M[j \to C_i] \cdot \pi_{C_i}^{\l}  \cdot x \right\vert\\
  & = \left\vert \frac{\W^{g_{\l}}_{L_m} (\o, \nu^*)}{L_m} - \left( \o(C_i) + \sum_{j \in T} \o^j \cdot M[j \to C_i] \right) \cdot \pi_{C_i}^{\l} \cdot x \right\vert,
\end{align*}
recovering the statement given in \eqref{Eq. nstar sufficient}, and thus concluding the proof of Theorem \ref{Uniform Value for MDP}.

\section{Proof of Theorem \ref{Thm Mono 1}}\label{Sec Proof of Mono 1}

We begin the proof by applying the dynamic programming principle for the MDP $\G_x$. Such a step will allow us to derive several preliminary results required for the proof of Theorem \ref{Thm Mono 1}. 

\subsection{The dynamic programming principle and its applications}

\subsubsection{The recursive formula in terms of the (Cav) operator}\label{Subsubsec Recursive Formula}

The dynamic programming principle applied to $\G_x$ gives rise, for each $N\geq 1$, to the following recursive formula
\begin{align}\label{Recursive 1}
    \Upsilon^x_{N} (\o) & = \sup_{\xi \in \Delta(A)^K} \bigg\lbrace \frac{1}{N} \cdot f(\o, \xi) + \frac{N-1}{N} \cdot x \cdot \sum_{\l \in K} \o^{\l} \cdot \, \Upsilon^x_{N-1}(\delta_{\l}^* M)\nonumber \\
    & \quad + \frac{N-1}{N} \cdot (1-x) \cdot \sum_{a \in A}\, \left[ \sum_{\l \in K} \o^{\l}\cdot \x^{\l}(a) \right] \, \Upsilon^x_{N-1} \left(\o(\x,a) M\right)  \bigg\rbrace.
\end{align}

We will now use the famous Aumann-Maschler Splitting Lemma \cite{Aumann}, to change the maximization domain in \eqref{Recursive 1}. Formally, this lemma states that for every sequence of beliefs $\{\rho_a \in \Delta(K)\,:\, a \in A \}$ and convex weights $\{\beta_a\,:\, a \in A\}$ satisfying $\o = \sum_{a \in A} \beta_a \rho_a $ there exists a lottery $\xi \in \Delta(A)^K$ such that 
\begin{align*}
    \o(\x,a) = \rho_a \quad \text{and}\quad  \sum_{\l \in K} \o^{\l}\cdot \x^{\l}(a) = \beta_a, \quad \forall a \in A.
\end{align*}

On the other hand, introducing the set of splits of $\o$, denoted $\mathcal{S}(\o)$, and defined by,
\begin{align*}
    \mathcal{S}(\o) := \bigg\lbrace (\beta_a, \rho_a)_{a \in A} \subset [0,1]\times \Delta(K) \,:\,  \sum_{a \in A} \beta_a = 1, \,\, \text{and}\,\,  \o = \sum_{a \in A} \beta_a \rho_a \bigg\rbrace
\end{align*}
we obtain that each action $\xi \in \Delta(A)^K$ defines the element $(\beta_a, \rho_a)_{a \in A}$ of $\mathcal{S}(\o)$ given by
\begin{align*}
    \beta_a := \sum_{\l \in K} \o^{\l}\cdot \x^{\l}(a) \quad \text{and} \quad \rho_a:=  \o(\x,a), \quad \forall a \in A. 
\end{align*}

Using the above correspondence between $\Delta(A)^K$ and $\mathcal{S}(\o)$, and reducing $f$ to the payoff associated with $MP(x)$, i.e.,  $f(\o,\x) = \sum_{a \in A} \left( \sum_{\l \in K} \o_n^{\l}\cdot \x_n^{\l}(a) \right) u(\o(\x,a))$, where $u$ is given \eqref{u-definition}, we may rewrite \eqref{Recursive 1} as follows:
\begin{align}\label{Recursive 2}
     \Upsilon^x_{N} (\o) & = \sup_{(\beta_a, \rho_a) \in \mathcal{S}(\o)} \bigg\lbrace \frac{1}{N}\sum_{a \in A} \beta_a \cdot u(\rho_a) + \frac{N-1}{N} \cdot x \cdot \sum_{\l \in K} \o^{\l} \cdot \, \Upsilon^x_{N-1}(\delta_{\l}^* M)\nonumber \\
    & \quad + \frac{N-1}{N} \cdot (1-x) \cdot \sum_{a \in A}\, \beta_a \, \Upsilon^x_{N-1} \left(\rho_a M\right)  \bigg\rbrace \nonumber\\
    & = \sup_{(\beta_a, \rho_a) \in \mathcal{S}(\o)} \bigg\lbrace \frac{1}{N}\sum_{a \in A} \beta_a \cdot u(\rho_a) + \frac{N-1}{N} \cdot (1-x) \cdot \sum_{a \in A}\, \beta_a \, \Upsilon^x_{N-1} \left(\rho_a M\right)  \bigg\rbrace \nonumber \\
    & \quad + \frac{N-1}{N} \cdot x \cdot \sum_{\l \in K} \o^{\l} \cdot \, \Upsilon^x_{N-1}(\delta_{\l}^* M)  \\
    & = \sup_{(\beta_a, \rho_a) \in \mathcal{S}(\o)} \sum_{a \in A}\, \beta_a \cdot\bigg\lbrace \frac{1}{N} \cdot u + \frac{N-1}{N} \cdot (1-x) \cdot \Upsilon^x_{N-1} \circ M  \bigg\rbrace (\rho_a) \nonumber \\
    & \quad + \frac{N-1}{N} \cdot x \cdot \sum_{\l \in K} \o^{\l} \cdot \, \Upsilon^x_{N-1}(\delta_{\l}^* M) \nonumber
\end{align}
where $\Upsilon^x_{N-1} \circ M$ denotes the concatenation of the $M$-shift $\o \mapsto \o M$ with $\Upsilon^x_{N-1}$. For each $N\geq 1$ introduce the function $\phi_N$ on $\Delta(K)$ defined by $\phi_N := \frac{1}{N} \cdot u + \frac{N-1}{N} \cdot (1-x) \cdot \Upsilon^x_{N-1} \circ M$. Carath\'{e}odory's Theorem (see, e.g., Corollary 17.1.5 in \cite{Rock}) implies that for any set $A$ satisfying $|A|\geq k$, one has 
\begin{align}\label{Eq. Caratheodory}
    \sup_{(\beta_a, \rho_a) \in \mathcal{S}(\o)} \sum_{a \in A}\, \beta_a \cdot\phi_N  (\rho_a)  = (\text{Cav}\, \phi_N) (\o), 
\end{align}

Next, let us introduce the affine function $L(\o) := \frac{N-1}{N} \cdot x \cdot \sum_{\l \in K} \o^{\l} \cdot \, \Upsilon^x_{N-1}(\delta_{\l}^* M)$. Combining the results of \eqref{Recursive 2} with \eqref{Eq. Caratheodory} we obtain that 
\begin{align}\label{Recursive 3}
\Upsilon^x_{N}   = (\text{Cav}\, \phi_N)+L
\end{align}

\subsubsection{The existence and properties of optimal strategies in $MP(x)$}\label{Subsubsec Optimal Strategies}

We begin with a functional analysis of the value $\Upsilon^x_{N}$. We have that $\Upsilon^x_{N}$ is a concave function on $\Delta(K)$  for every $N$, as by Eq.\ \eqref{Recursive 3} it is the sum of two concave functions. 
The concavity of $\Upsilon^x_{N}$ is of importance as it implies that $\phi_N$ is an upper semicontinuous function, being a weighted sum of the upper semicontinuous function $u$ with the continuous (concave) function $\Upsilon^x_{N-1} \circ M$. In such a case, it is known that the supremum in \eqref{Eq. Caratheodory} is attained. Moreover, Carath\'{e}odory's Theorem (see, e.g., Corollary 17.1.5 in \cite{Rock}) implies that for each $\o \in \Delta(K)$, this supremum can be attained by a split to at most $k$ elements (i.e., $\{(\b_i,\rho_i)\,:\, i = 1,...,k\} \in \mathcal{S}(\o)$).

For the sake of the current and subsequent discussions, let us fix $N\geq 1$. The strategic implication of the preceding paragraph coupled with \eqref{Recursive 2} provides an optimal \textbf{pure} strategy $\nu^{x} \in \mathcal{N}$, that takes at each stage $n$, given the state $\o_n = \chi$, an action $\xi_n^{x}(\chi)$ that satisfies
\begin{align*}
    \chi(\xi_n^{x}(\chi), \cdot) = \rho_i^{x,n} (\chi) \quad \text{with probability} \quad \beta_i^{x,n} (\chi),
\end{align*}
where 
\begin{align*}
    (\text{Cav}\, \phi_{N-n+1}) (\chi) = \sum_{i=1}^k \, \beta_i^{x,n} (\chi) \cdot\phi_{N-n+1} (\rho_i^{x,n} (\chi)). 
\end{align*}

The strategy $\nu^x$ may be implemented by the sender in $MP(x)$, as we assumed that the signal set $S$ contains at least $k$ signals. Thus, as the analysis above was carried for $\Gamma_x$ with the reduced payoff $f$ corresponding to the payoff in $MP(x)$ (e.g., Subsection \ref{MDP of MP(x)}), we learn that the values $v_N^x$ of $MP(x)$ remain the same for all signal sets $S$ satisfying $|S| \geq k$. In particular, there is no additional benefit in taking $A$ to be countable as opposed to finite (as assumed for $S$ in $MP(x)$).

We may therefore continue working with the MDP $\G_x$ with the reduced $f$, and general signal set $A$, and obtain that any result that holds for the value $\Upsilon^x_{N}$ is also applicable to the values $v_N^x$, as by the preceding paragraph, both are the same. 

\subsection{The Monotonicity of $\Upsilon^x_{N}$ in $x$}\label{Subsection Monotonicity Proof}

Keeping in mind that we fixed $N\geq 1$, we now state the following key proposition.
\begin{proposition}\label{Lemma Monotonicity of V-N}
    The mapping $x \mapsto \Upsilon^x_{N}(\o)$ is non-increasing on $(0,1]$ for every $\o \in \Delta(K)$.
\end{proposition}

\begin{proof}[Proof of Proposition \ref{Lemma Monotonicity of V-N}]

The proof strategy is the following. Let us fix $0<x<y\leq 1$. We have that $\Upsilon^y_N (\o) = \U^x_N (\o, \nu^{y})$, where $\nu^{y}$ was introduced in Subsection \ref{Subsubsec Optimal Strategies}. Therefore, a construction of a strategy $\nu^* \in \mathcal{N}$ such that $\U^x_N (\o, \nu^*) = \U^y_N (\o, \nu^{y})$ will imply that $\Upsilon^y_N (\o) \leq \Upsilon^x_N (\o)$, thus proving the proposition.

To construct $\nu^*$, we first associate with each $\o \in \Delta(K)$ and $n=1,...,N$, an action $\zeta^*_n (\o) \in \Delta(A)^K$ which induces the following split in $\mathcal{S}(\o M)$:

\begin{align}
    (\o M)(\zeta^*_n(\o), \cdot) = \rho_i^{y,n} (\d_{\l} M)  \quad \text{with probability} \quad  \o^{\l} \cdot \beta_i^{y,n}(\d_{\l} M).
\end{align}
In words, $\zeta^*_n (\o)$ performs a `double split' of $\o M$. First it splits $\o M$ to the shifted Dirac measures $\{\d_{\l}M\}$ according to $\{\o^{\l}\}$ respectively, and then splits each $\d_{\l}M$ according to the action $\xi_n^y (\d_{\l}M)$ described in Subsection \ref{Subsubsec Optimal Strategies}. 

We are now in a position to define $\nu^*$. The strategy $\nu^*$ is defined in terms of the sequence of actions $(\xi^*_n)_{n\geq 1}$ defined iteratively by:
\begin{align*}
     \xi^*_1 (\o)  := \xi^y_1 (\o),
\end{align*} 
    and for each $n\geq 2$,
\[
 \xi^*_n (\o, \xi^*_1,..., \o_{n-1}, \xi^*_{n-1}, \o_n) = 
  \begin{cases} 
   \xi_n^y (\d_{\l}M) & \text{if }\o_n = \d_{\l}^*M,\, \l \in K \\
   \xi_n^y (\o_n)       & \text{with prob.\  }\,\, \frac{1-y}{1-x}\,\, \text{if}\,\, \o_n \notin \R\\
   \zeta^*_n \left( \o_{n-1}(\xi^*_{n-1}, a) \right) & \text{with prob.\  }\,\, \frac{y-x}{1-x}\,\, \text{if}\\
   & \,\, \o_n  = \o_{n-1}(\xi^*_{n-1}, a) M,\,\, a \in A.  \end{cases}
  \]

\begin{lemma}\label{Lemma Distribution Equality}
    For every $n=1,...,N$ it holds that 
    \begin{align*}
        P^{\o}_{x,\nu^*} \left[ \o_n(\xi_n^*, \cdot) \in B  \right] = P^{\o}_{y,\nu^y} \left[ \o_n(\xi_n^y, \cdot) \in B  \right], \quad \forall B \in \mathcal{B}(\Delta(K)).
    \end{align*}
    That is, the marginal distribution of posteriors in $\Gamma_x$ under $\nu^*$ identifies with the marginal distribution of posteriors in $\Gamma_y$ under $\nu^y$.
\end{lemma}
\begin{proof}[Proof of Lemma \ref{Lemma Distribution Equality}]
   To simplify the calculations in the proof, we distinguish between elements in the support of $\o(\xi,\cdot)$ and $\o'(\xi,\cdot)$ for any two states $\o,\o'$ and action $\xi$. That is, even if some belief has positive probability under both $\o(\xi,\cdot)$ and $\o'(\xi,\cdot)$, we view this belief as having two distinguished copies coming from the latter two distributions. This distinction has no effect on the total probabilities of elements in the supports of  $\o_n(\xi_n^*, \cdot)$ and $\o_n(\xi_n^y, \cdot)$, and is therefore compatible.

   The proof proceeds by induction on $n$. The base case $n=1$ follows immediately from the definition of $\nu^*$. For the induction step, assume that the result holds for $n-1$ and let us prove it for $n$. The proof goes in parts. 
   
   \begin{claim}\label{Claim Step 1}
   $\o_n(\xi_n^*, \cdot)$ and $\o_n(\xi_n^y, \cdot)$ have the same \textbf{conditional} distribution on $\{\o_n \in \R\}$.
   \end{claim}

   \begin{proof}[Proof of Claim \ref{Claim Step 1}]
   The definition of $\xi_n^*$ implies that $\o_n(\xi_n^*, \cdot)$ and $\o_n(\xi_n^y, \cdot)$ have the same conditional distribution on each event $\{\o_n = \d_{\l}^* M\}$, $\l \in K$. Therefore, the claim follows from the fact that for any $t \in (0,1]$ and $\nu \in \mathcal{N}$ one has
   \begin{align*}
       P^{\o}_{t,\nu} ( \{\o_n = \d_{\l}^* M\}\,|\, \{\o_n \in \R\} ) & = E^{\o}_{t,\nu}\left( P^{\o}_{t,\nu} ( \{\o_n = \d_{\l}^* M\}\,|\, \{\o_n \in \R\}, \o_{n-1})\right)\\
       & = \frac{1}{t} E^{\o}_{t,\nu}\left( P^{\o}_{t,\nu} ( \{\o_n = \d_{\l}^* M\})\,|\, \o_{n-1})\right)\\
       & = \frac{1}{t} E^{\o}_{t,\nu}( t \cdot \o_{n-1}^{\l})\\
       & = ( \o M^{n-2} )^{\l}.
   \end{align*}
   \end{proof}

   \begin{claim}\label{Claim Step 2}
   $\o_n(\xi_n^*, \cdot)$ has the same \textbf{conditional} distribution on both $\{\o_n \in \R\}$ and $\{\o_n \notin \R,\, \xi_n^* \neq \xi_n^y (\o_n)\}$.
   \end{claim}

\begin{proof}[Proof of Claim \ref{Claim Step 2}]
   Let $\chi$ be an element in the support of $(p_{\l})(\xi_n^y,\cdot)$ for some $\l \in K$. That is, $\chi =\rho_i^{y,n} (\d_{\l} M)$ for some $i$. We have that
   \begin{align*}
       & P^{\o}_{x,\nu^*} \left( \o_n(\xi_n^*, \cdot) = \chi\,|\, \{\o_n \notin \R,\, \xi_n^* \neq \xi_n^y (\o_n)\} \right)\\
       & = E^{\o}_{x,\nu^*}\left( P^{\o}_{x,\nu^*} \left( \o_n(\xi_n^*, \cdot) = \chi\,|\, \{\o_n \notin \R,\, \xi_n^* \neq \xi_n^y (\o_n)\}, \o_{n-1}(\xi^*_{n-1}, \cdot) \right)  \right)\\
       & = E^{\o}_{x,\nu^*}\left( (\o_{n-1}(\xi^*_{n-1}, \cdot))^{\l} \cdot \beta_i^{y,n}(\d_{\l} M)  \right)\\
       & = \beta_i^{y,n}(\d_{\l} M) \cdot \left(E^{\o}_{x,\nu^*}(\o_{n-1}(\xi^*_{n-1}, \cdot))\right)^{\l}\\
       & = \beta_i^{y,n}(\d_{\l} M) \cdot \left(E^{\o}_{x,\nu^*}\left(\o_{n-1}\right)\right)^{\l} = \beta_i^{y,n}(\d_{\l} M) \cdot ( \o M^{n-2} )^{\l},
   \end{align*}
   where the third equality follows from the fact that the mapping $\o \mapsto \o^{\l}$ is affine, and the fourth equality follows from the fact that given $\o_{n-1}$, $\o_{n-1}(\xi^*_{n-1}, \cdot)$ has mean $\o_{n-1}$.
   The proof of the claim is completed, since by arguments parallel to those given in the proof of Claim \ref{Claim Step 1} we have $P^{\o}_{x,\nu^*} \left( \o_n(\xi_n^*, \cdot) = \chi\,|\, \{\o_n \in \R\} \right) = \beta_i^{y,n}(\d_{\l} M) \cdot ( \o M^{n-2} )^{\l}$ as well. 
   \end{proof}

    \begin{claim}\label{Claim Step 3}
   The (unconditional) distribution of $\o_n(\xi_n^*, \cdot)$ on $\{\o_n \in \R\}$ and $\{\o_n \notin \R,\, \xi_n^* \neq \xi_n^y (\o_n)\}$ agrees with the (unconditional) distribution of $\o_n(\xi_n^y, \cdot)$ on $\{\o_n \in \R\}$.
   \end{claim}
  
\begin{proof}[Proof of Claim \ref{Claim Step 3}]
    The result is an immediate consequence of Claims \ref{Claim Step 1} and \ref{Claim Step 2} together with the fact that $P^{\o}_{y,\nu^y}(\{\o_n \in \R\}) = y$ and
    \begin{align}\label{Eq. Coup. Mon.}
        P^{\o}_{x,\nu^*}\left( \{\o_n \in \R\} \right) + P^{\o}_{x,\nu^*}\left( \{\o_n \notin \R,\, \xi_n^* \neq \xi_n^y (\o_n)\} \right) = x + (1-x)\cdot \frac{y-x}{1-x} = y.
    \end{align}
\end{proof}

In view of Claim \ref{Claim Step 3}, to prove the lemma, it suffices to show the following claim.

    \begin{claim}\label{Claim Step 4}
   The conditional distribution of $\o_n(\xi_n^*, \cdot)$ on $\{\o_n \notin \R,\, \xi_n^* = \xi_n^y (\o_n)\}$ agrees with the conditional distribution of $\o_n(\xi_n^y, \cdot)$ on $\{\o_n \notin \R\}$.
   \end{claim}

\begin{proof}[Proof of Claim \ref{Claim Step 4}]
 Let $\chi$ be a belief in $\Delta(K)$. We have
 \begin{align*}
& P^{\o}_{x,\nu^*}\left( \o_n(\xi_n^*, \cdot) = \chi\,|\, \{\o_n \notin \R,\, \xi_n^* = \xi_n^y (\o_n)\}  \right)\\
& = E^{\o}_{x,\nu^*}\left( P^{\o}_{x,\nu^*}\left( \o_n(\xi_n^*, \cdot) = \chi \,|\, \{\o_n \notin \R,\, \xi_n^* = \xi_n^y (\o_n)\},\, \o_{n-1}(\xi_{n-1}^*, \cdot) = q  \right)
\right)\\
& = \frac{1}{1-y} \cdot E^{\o}_{x,\nu^*}\left( P^{\o}_{x,\nu^*}\left( (qM)(\xi_n^y (qM), \cdot) = \chi \,|\, \o_{n-1}(\xi_{n-1}^*, \cdot)= q  \right)
\right)\\
& = \frac{1}{1-y} \cdot P^{\o}_{x,\nu^*}\left( (\o_{n-1}(\xi_{n-1}^*, \cdot)M)(\xi_n^y (\o_{n-1}(\xi_{n-1}^*, \cdot)M), \cdot) = \chi\right)\\
& = \frac{1}{1-y} \cdot P^{\o}_{y,\nu^y}\left( (\o_{n-1}(\xi_{n-1}^y, \cdot)M)(\xi_n^y (\o_{n-1}(\xi_{n-1}^y, \cdot)M), \cdot) = \chi\right)\\
& = P^{\o}_{y,\nu^y}\left( \o_n(\xi_n^y, \cdot) = \chi\,|\, \{\o_n \notin \R\}  \right),
 \end{align*}
 where the second inequality follows from the definition of $\xi_n^*$ and Eq.\ \eqref{Eq. Coup. Mon.}, the third from the law of total expectation, the fourth is due to the induction hypothesis, and the last one is yet another consequence of the law of total expectation.
\end{proof}
\end{proof}

We are now in a position to deduce the proposition. Indeed, under the reduced payoff function $f$, the payoffs $\U^x_N (\o, \nu^*)$ and $\U^y_N (\o, \nu^{y})$ are determined only by the marginal distributions of the random variables $\{\o_n(\xi_n^*, \cdot)\,:\,n=1,...,N\}$ and $\{\o_n(\xi_n^y, \cdot)\,:\, n=1,...,N\}$, respectively. Hence, by Lemma \ref{Lemma Distribution Equality} we obtain that $\U^x_N (\o, \nu^*)= \U^y_N (\o, \nu^{y})$. As discussed at the start of the proof, this is sufficient to deduce that $\Upsilon^x_N (\o) \geq \Upsilon^y_N(\o)$ for each $N \geq 1$.
\end{proof}

We conclude the section with the proof of Theorem \ref{Thm Mono 1}.

\begin{proof}[Proof of Theorem \ref{Thm Mono 1}.]
We saw in the proof of Theorem \ref{Uniform Value for MDP} that $\Upsilon^x_N (\cdot)$ converges uniformly to $\V^*(\cdot,x)$ as $N \to \infty$, which for an irreducible $M$ equals to the constant $\pi_M^1 \Upsilon_{1-x}(\textbf{m}_1) + \cdots + \pi_M^k \Upsilon_{1-x}(\textbf{m}_k)$. Therefore, by Proposition \ref{Lemma Monotonicity of V-N}, we deduce that the mapping 
\begin{align*}
    x \mapsto \pi_M^1 \Upsilon_{1-x}(\textbf{m}_1) + \cdots + \pi_M^k \Upsilon_{1-x}(\textbf{m}_k)
\end{align*}
is non-increasing in $(0,1]$. Taking $\delta(x) = 1-x$, we obtain that the mapping 
\begin{align*}
    \delta \mapsto \pi^1 \Upsilon_{\delta}(\textbf{m}_1) + \cdots + \pi^k \Upsilon_{\delta}(\textbf{m}_k)
\end{align*}
is non-decreasing on $[0,1)$, thus proving Theorem \ref{Thm Mono 1}.
\end{proof}

\section{Proof of Theorems \ref{Thm Mono 2} and \ref{Thm RG-mono}}\label{sec: proof mono 2}

\subsection{Proof of Theorem \ref{Thm Mono 2}}\label{SubSec Proof Mono 2}

Our proof for Theorem \ref{Thm Mono 2} is accomplished by introducing a suitable generalization of the MDP $\G_x$ defined in Section \ref{Section General MDP}, and thereafter performing a close analysis to that given in Sections \ref{Proof of Uniform and Asym Values} and \ref{Sec Proof of Mono 1}, with minor adjustments where needed.   

Our starting point is to define $C = conv \{\chi_1,...,\chi_k\}$ and note that by assumption, for every $\o \in C$ there exist unique convex weights $\gamma_1 (\o),..., \gamma_k (\o)$ such that 
\begin{align*}
    \o = \gamma_1 (\o) \cdot \chi_1 + ... + \gamma_k (\o) \cdot \chi_k.
\end{align*}
Moreover, by the uniqueness of the convex decomposition, we have that the mapping $\gamma_i : \o \mapsto \gamma_i (\o)$ is an affine one for every $i \in \{1,...,k\}$, and in particular continuous.

The generalization of the MDP $\G_x$ we shall introduce, will be denoted by $\G_x^*$, and described by the following rules:
\begin{itemize}
    \item The \textit{states} of $\Gamma_x^*$ are $\Delta(K) \cup \R$, where $\R := \{\chi^*_1,...,\chi^*_k \}$ is a distinguished copy of the beliefs $\{\chi_1,...,\chi_k\}$.
    \item The \textit{actions} in $\Gamma_x^*$ are the same as in $\Gamma_x$, being  lotteries $\x$ over a countable set $A$ which may depend on $K$, i.e., elements in $\D(A)^K$.   
    \item The \textit{transition rule}, denoted $t^*$, is defined by
\begin{equation*}
t^*(\o,\x) = \left\{
       \begin{array}{ll}
        \chi_{\l}^* , & \hbox{with prob.\,\,}\,\, x\cdot \gamma_{\l}(\o M),	\,\,\,\, \l \in K,\\ 
        \o(\x,a) M , & \hbox{with prob.\,\,}\,\, (1-x)\sum_{\l \in K} \o^{\l}\cdot \x^{\l}(a),	\,\,\,\, a \in A,
       \end{array}
     \right.
\end{equation*}
    \item The \textit{payoff} $f(\o,\x)$ associated with state $\o \in \Delta(K) \cup \R^*$ and action $\xi \in \D(A)^K$ is given by $f(\o,\x) = \sum_{a \in A} \left( \sum_{\l \in K} \o_n^{\l}\cdot \x_n^{\l}(a) \right) u(\o(\x,a))$, where $u$ is given \eqref{u-definition}.
\end{itemize}

The MDP $\G_x^*$ is well defined since $\o M \in C$ for every $\o \in \dk \cup \R^*$.\footnote{For $\o = \chi_{\l}^*$, $\l = 1,...,k$, we define $\o M = \chi_{\l}M$.} As the space of behavioral strategies in $\Gamma_x^*$ agrees with that in $\Gamma_x$, we shall use the same notations for the $N$-Ces\`aro and $\delta$-discounted payoffs and values in $\Gamma_x^*$ as we did in $\Gamma_x$, whenever $x>0$. Moreover, we note that in the case where $x = 0$, the transition rules in both $\Gamma_x^*$ and  $\Gamma_x$ identify, so that for the payoff $f$ specified above, the $N$-Ces\`aro and $\delta$-discounted values are the same for both $\G_0$ and $\G_0^*$ starting from any $\o \in C$. 

\begin{remark}\label{Remark 2}
    Note that the MDP $\G_x^*$ can be reduced to a modified version of $MP(x)$ in which at each stage $n\geq 1$, the stochastic revelation occurs before the receiver chooses an action.\footnote{In such a scenario, the sender's signal $s_n$ does not affect the receiver's belief.} Indeed, this reduction is achieved by choosing $\chi_{\l} = \d_{\l}$ for every $\l \in K$ and $A=S$.\footnote{Note that the choice $A=S$ is without loss of generality as follows from arguments similar to those given in Subsection \ref{Subsubsec Recursive Formula}, i.e., combining Carath\'{e}odory's Theorem with the respective recursive formulas for $\G_x^*$.} Thus, all preceding results for such a choice provide intuition as to how the advancement of the stochastic revelation within the timeline of each stage (e.g., Section \ref{sec: model}) affects the results regarding $MP(x)$.
\end{remark}

Let us argue that to prove Theorem \ref{Thm Mono 2} it suffices to establish the following three propositions:
\begin{proposition}\label{Prop Mono 2-1}
    For every $x \in (0,1]$, $\Upsilon^x_N (\pi_M)$ converges to $\g_1(\pi_M) \cdot \Upsilon_{1-x}(\chi_1) + ... + \g_k(\pi_M) \cdot \Upsilon_{1-x}(\chi_k)$ as $N \to \infty$.
\end{proposition}

\begin{proposition}\label{Prop Mono 2-2}
      The mapping $x \mapsto \Upsilon^x_{N}(\pi_M)$ is non-increasing on $(0,1]$ for every $\o \in C$.
\end{proposition}

\begin{proposition}\label{Prop Mono 2-3}
     $\Upsilon_{\delta}(\o) = v_{\delta}(\o)$ for every $\o \in \Delta(K)$.
\end{proposition}

Indeed, the combination of Propositions \ref{Prop Mono 2-1} and \ref{Prop Mono 2-2} implies that $ x \mapsto \g_1(\pi_M) \cdot \Upsilon_{1-x}(\chi_1) + ... + \g_k(\pi_M) \cdot \Upsilon_{1-x}(\chi_k)$ is non-increasing on $(0,1]$. Therefore, by Proposition \ref{Prop Mono 2-3} the mapping $ x \mapsto \g_1(\pi_M) \cdot v_{1-x}(\chi_1) + ... + \g_k(\pi_M) \cdot v_{1-x}(\chi_k)$ is non-increasing on $(0,1]$. Therefore, the mapping $ \delta \mapsto \g_1(\pi_M) \cdot v_{\delta}(\chi_1) + ... + \g_k(\pi_M) \cdot v_{\delta}(\chi_k)$ is non-decreasing on $[0,1)$, as required.
\bigskip

Before proceeding with the proofs, let us discuss the so-called `mean-consistency' property at $\pi_M$ of the MDP $\Gamma_x^*$, which will play a key role in the analysis. Formally, using an inductive argument, one can verify that the following holds for $\Gamma_x^*$, just as in $\Gamma_x$:  
\[
E^{\o}_{x,\nu} \,\o_n = \o M^{n-1},
\]
for every $\o$, $\nu \in \mathcal{N}$, and $n \geq 1$. In particular, taking $\o = \p$, we obtain the following consistency of means at $\p$:
\[
E^{\pi_M}_{x,\nu} \o_n = \pi_M, \quad \forall n\geq 1,\,\,\, \forall \nu \in \mathcal{N}.
\]  

We move on with the proof of Proposition \ref{Prop Mono 2-1}.

\subsubsection{Proof of Proposition \ref{Prop Mono 2-1}}\label{Proof of Mono 2 Convergence}

The starting point of the proof is the adoption of the same probabilistic approach for $\G_x^*$, as the one described for $\G_x$ in Subsection \ref{Subsec Prob Approach}. 
In particular, using the same notations, we obtain that Lemma \ref{Lemma Approx. 1} and Claim \ref{Claim Approx. 1} are also valid for the current setup.

Next, we define the appropriate generalization of the strategy $\nu^* \in \mathcal{N}$ (previously defined in Subsection \ref{Subsec nu-star}) as follows:
\begin{itemize}
    \item Play arbitrary until stage $T_1 + 1$.
    \item For every $1 \leq m$, starting from stage $T_m+1$, conditional on $\o_{T_m + 1} = \chi_{\l}^*$, follow the strategy $\hat{\nu}_{1-x}(\chi_{\l})$, until stage $T_{m+1}$ (included).
\end{itemize}
By the exact same arguments described in Subsection \ref{Subsec nu-star optimality} we obtain that 
\begin{align}\label{Eq. Optimality of nu* 1 *}
    E_{x,\nu}^{\pi_M} \left( \sum_{n = T_i+1}^{ T_{i+1}} f(\o_n,\x_n)\,|\, T_i +1 , \o_{T_i +1}  \right) \leq  E_{x,\nu^*}^{\pi_M} \left( \sum_{n = T_i+1}^{ T_{i+1}} f(\o_n,\x_n)\,|\, T_i +1 , \o_{T_i +1}  \right),
\end{align}
for every $i\geq 1$ and every $\nu \in \mathcal{N}$. By the same arguments described in Subsection \ref{Subsec nu-star optimality}, we have that for every $i\geq 1$ the marginal distribution of $(T_i+1 , \o_{T_i+1})$ is independent of $\nu \in \mathcal{N}$. 
Next, for every $\l \in K$ and $\nu \in \mathcal{N}$ we have
\begin{align}\label{Eq. marginal state expectation *}
& P_{x,\nu}^{\pi_M} \left(\o_{T_i+1} = \chi_{\l}^*\, |\, T_i+1 = r \right) =  E_{x,\nu}^{\pi_M} \left( \1\{\o_{T_i+1} = \chi_{\l}^* \} \,|\, T_i+1 = r \right) \nonumber \\
   & \quad \quad \quad \quad \quad \quad =  E_{x,\nu}^{\pi_M} \left( E_{x,\nu}^{\pi_M} \left( \1\{\o_{T_i+1} = \chi_{\l}^* \} \,|\, T_i+1 = r, \o_{r-1} \right) \right)\\
   & \quad \quad \quad \quad \quad \quad = E_{x,\nu}^{\pi_M} \left(\gamma_{\l}(\o_{r-1} M)\right) = \gamma_{\l}\left( (E_{x,\nu}^{\pi_M}\, \o_{r-1} )M \right) = \gamma_{\l}(\pi_M), \nonumber
\end{align}
where the penultimate equality follows from the affinity of $\gamma_{\l}$ and the $M$-shift, whereas the last equality follows from the mean-consistency property at $\pi_M$. 

Therefore, we may use the law of total expectation on both sides in \eqref{Eq. Optimality of nu* 1 *} and sum over $i=1,...,m-1$ to obtain
\begin{align*}
    \mathcal{Q}_m(\o,\nu) \leq \mathcal{Q}_m(\o,\nu^*), \quad \forall \nu \in \mathcal{N}.
\end{align*}
which, in turn, together with Claim \ref{Claim Approx. 1}, implies that 
\begin{align}\label{Eq. Two sided in G* 1}
    \left\vert \Upsilon_{L_m}^x(\pi_M) - \frac{1}{L_m} \mathcal{Q}_m(\o,\nu^*) \right\vert \leq  o_m(1).
\end{align}
Next, by performing an analogous analysis to that in Subsection \ref{Subsec nu-star payoff}, we obtain the parallel result to that of Eq.\ \eqref{Eq. General Mono Approx.}, in the form of
\begin{align}\label{Eq. Two sided in G* 2}
    \left\vert  \frac{1}{L_m} \mathcal{Q}_m (\o, \nu^*) -  \sum_{\l \in K} \frac{\W^{g_{\l}}_{L_m} (\o, \nu^*)}{L_m} \cdot \Upsilon_{1-x} (\chi_{\l}) /x    \right\vert \leq o_m(1).
\end{align}
where in the current setup we let $g_{\l}(\o,\xi) := \1\{ \o = \chi_{\l}^* \}$ for every $(\o,\xi)$. Using this definition, we have that 
\begin{align*}
     \frac{\W^{g_{\l}}_{L_m} (\pi_M, \nu^*)}{L_m} & = E_{x,\nu^*}^{\pi_M} \left( \frac{1}{L_m}\sum_{n=2}^{L_m}  \1\{\o_n = \chi_{\l}^* \} \right)\\
     & = \frac{1}{L_m} \sum_{n=2}^{L_m} E_{x,\nu^*}^{\pi_M} \left( E_{x,\nu^*}^{\pi_M} ( \1\{\o_n = \chi_{\l}^* \}\,|\, \o_{n-1} ) \right) \nonumber \\
    & = \frac{1}{L_m} \sum_{n=2}^{L_m} E_{x,\nu^*}^{\pi_M} \left(x \cdot \gamma_{\l}(\o_{n-1}M) \right) \\
    & = \frac{1}{L_m} \sum_{n=2}^{L_m}  x \cdot \gamma_{\l}\left( \left[ E_{x,\nu^*}^{\pi_M}(\o_{n-1}) \right]  M \right) \\
    & = \frac{1}{L_m} \sum_{n=2}^{L_m}  x \cdot \gamma_{\l}(\pi_M  M) = \frac{L_m-1}{L_m} \cdot x \cdot \gamma_{\l}(\pi_M),
\end{align*}
where the third equality follows from the definition of $\Gamma_x^*$, the fourth from the affinity of both $\gamma_{\l}$ and the $M$-shift operator, and the penultimate equality from the mean-consistency property at $\pi_M$. 

The above derivation implies that $\left\vert \W^{g_{\l}}_{L_m} (\pi_M, \nu^*)/L_m -  x \cdot \gamma_{\l}(\pi_M) \right\vert \leq o_m(1)$, which together with Eqs.\ \eqref{Eq. Two sided in G* 1}, \eqref{Eq. Two sided in G* 2}, and the fact $\left\vert \Upsilon_{1-x} (\cdot) /x \right\vert \leq \Vert f \Vert/x$ implies that $\Upsilon_{L_m}^x(\pi_M)$ converges to $\g_1(\pi_M) \cdot \Upsilon_{1-x}(\chi_1) + ... + \g_k(\pi_M) \cdot \Upsilon_{1-x}(\chi_k)$ as $m \to \infty$. This, in conjunction with the application of the bounded increments property of $(L_m)_{m \geq 1}$ given in Eq.\ \eqref{Eq. bounded increments app} implies that the sequence $(\Upsilon_{N}^x(\pi_M))_{N\geq 1}$ must converge to the same limit as well, thus completing the proof of the proposition.
\bigskip

We move on with the proof of Proposition \ref{Prop Mono 2-2}.
\subsubsection{Proof of Proposition \ref{Prop Mono 2-2}}\label{Proof of Mono 2 Monotonicity}
We begin the proof with a similar recursive formula approach, as we did in Subsection \ref{Subsubsec Recursive Formula}. Following parallel arguments, we obtain that for a fixed $N\geq 1$ it holds
\begin{align*}
     \Upsilon^x_{N} (\o) & = \sup_{(\beta_a, \rho_a) \in \mathcal{S}(\o)} \sum_{a \in A}\, \beta_a \cdot\bigg\lbrace \frac{1}{N} \cdot u + \frac{N-1}{N} \cdot (1-x) \cdot \Upsilon^x_{N-1} \circ M  \bigg\rbrace (\rho_a) \nonumber \\
    & \quad \quad \quad + \frac{N-1}{N} \cdot x \cdot \sum_{\l \in K} \gamma_{\l}(\o M) \cdot \, \Upsilon^x_{N-1}(\chi_{\l}^*)  \\
    & = (\text{Cav}\, \phi_N)(\o)+L^*(\o) \nonumber,
\end{align*}
for any $\o \in C$ and any $x \in [0,1]$, where $\phi_N : \Delta(K) \to \mathbb{R}$ is defined by $ \phi_N (\o) := \frac{1}{N} \cdot u (\o) + \frac{N-1}{N} \cdot (1-x) \cdot (\Upsilon^x_{N-1} \circ M)(\o)$, and $L^* : \Delta(K) \to \mathbb{R}$ is defined by $L^*(\o) := \frac{N-1}{N} \cdot x \cdot \sum_{\l \in K} \Upsilon^x_{N-1}(\chi_{\l}^*) \cdot (\gamma_{\l}\circ M) (\o)$.

The mapping $L^*$ is the sum of affine functions and thus is continuous. Therefore, $\Upsilon^x_{N}$ is continuous on $\Delta(K)$ for every $N$, being the sum of $L^*$ with the concave function $(\text{Cav}\, \phi_N)$. Therefore, $\phi_N$ is upper semi-continuous, being the weighted average of the upper semicontinuous $u$ with the continuous function $\Upsilon^x_{N-1} \circ M$. Thus, for the fixed $N\geq 1$, by the same lines of arguments as in Subsection \ref{Subsubsec Optimal Strategies}, we obtain an optimal \textbf{pure} strategy $\nu^{x} \in \mathcal{N}$, that takes at each stage $n$, given the state $\o_n = \chi$, an action $\xi_n^{x}(\chi)$ that satisfies
\begin{align*}
    \chi(\xi_n^{x}(\chi), \cdot) = \rho_i^{x,n} (\chi) \quad \text{with probability} \quad \beta_i^{x,n} (\chi),
\end{align*}
where 
\begin{align*}
    (\text{Cav}\, \phi_{N-n+1}) (\chi) = \sum_{i=1}^k \, \beta_i^{x,n} (\chi) \cdot\phi_{N-n+1} (\rho_i^{x,n} (\chi)). 
\end{align*}

As $\Upsilon^y_N (\pi_M) = \U^y_N (\pi_M, \nu^{y})$ for every $y \in (0,1]$, to prove Proposition \ref{Prop Mono 2-2} it suffices to show that for every $0<x<y$ there exists a strategy $\nu^* \in \mathcal{N}$ for which $\U^y_N (\pi_M, \nu^{y}) = \U^x_N (\pi_M, \nu^*)$. The strategy $\nu^*$ will be the natural generalization of the strategy $\nu^*$ defined in Subsection \ref{Subsection Monotonicity Proof}. That is, the only change in the definition of $\nu^*$ compared to that in Subsection \ref{Subsection Monotonicity Proof} will arise by adjusting the definition of the action $\zeta^*_n (\o) \in \Delta(A)^K$ to induce the following split in $\mathcal{S}(\o M)$:

\begin{align}
    (\o M)(\zeta^*_n(\o), \cdot) = \rho_i^{y,n} (\chi_{\l})  \quad \text{with probability} \quad  \gamma_{\l}(\o M) \cdot \beta_i^{y,n}(\chi_{\l}).
\end{align}
for every $\o \in \Delta(K)$ and $n=1,...,N$. That is, in the current generalization, $\zeta^*_n (\o)$ performs a `double split' of $\o M$, first splitting it to the beliefs $\{\chi_{\l}\}$ according to $\{\gamma_{l}(\o M)\}$ respectively, and then splitting each $\chi_{\l}$ according to the action $\xi_n^y(\chi_{\l})$ described above. 

Thereafter, the proof that $\U^y_N (\pi_M, \nu^{y}) = \U^x_N (\pi_M, \nu^*)$ follows by proving Lemma \ref{Lemma Distribution Equality} just for the prior $\o = \pi_M$, 
under the current framework. Such a proof is carried out by the same sequence of claims, with some small technical adaptations that the current setup requires. We now discuss those adaptations in detail.

 \begin{proof}[Proof of Claim \ref{Claim Step 1}]
   The definition of $\xi_n^*$ implies that $\o_n(\xi_n^*, \cdot)$ and $\o_n(\xi_n^y, \cdot)$ have the same conditional distribution on each event $\{\o_n = \chi_{\l}^* \}$, $\l \in K$. Therefore, the claim follows from the fact that for any $t \in (0,1]$ and $\nu \in \mathcal{N}$ one has
   \begin{align*}
       P^{\pi_M}_{t,\nu} ( \{\o_n = \chi_{\l}^*\}\,|\, \{\o_n \in \R\} ) & = E^{\pi_M}_{t,\nu}\left( P^{\o}_{t,\nu} ( \{\o_n = \chi_{\l}^*\}\,|\, \{\o_n \in \R\}, \o_{n-1})\right)\\
       & = \frac{1}{t} E^{\pi_M}_{t,\nu}\left( P^{\o}_{t,\nu} ( \{\o_n = \chi_{\l}^*\})\,|\, \o_{n-1})\right)\\
       & = \frac{1}{t} E^{\pi_M}_{t,\nu}\left( t \cdot \gamma_{\l}(\o_{n-1} M) \right)\\
       & = \gamma_{\l} \left( E^{\pi_M}_{t,\nu}\, \o_{n-1} M \right) \\
       & = \gamma_{\l}(\pi_M) 
   \end{align*}
   where the previous to last equality is due to the affinity of $\gamma_{\l}$, and the last one is due to the `mean-consistency' property at $\pi_M$, and the affinity of the $M$-shift.
   \end{proof}

\begin{proof}[Proof of Claim \ref{Claim Step 2}]
   Let $\chi$ be an element in the support of $\chi_{\l}(\xi_n^y,\cdot)$ for some $\l \in K$. That is, $\chi =\rho_i^{y,n} (\chi_{\l})$ for some $i$. We have that
   \begin{align*}
       & P^{\pi_M}_{x,\nu^*} \left( \o_n(\xi_n^*, \cdot) = \chi\,|\, \{\o_n \notin \R,\, \xi_n^* \neq \xi_n^y (\o_n)\} \right)\\
       & = E^{\pi_M}_{x,\nu^*}\left( P^{\pi_M}_{x,\nu^*} \left( \o_n(\xi_n^*, \cdot) = \chi\,|\, \{\o_n \notin \R,\, \xi_n^* \neq \xi_n^y (\o_n)\}, \o_{n-1}(\xi^*_{n-1}, \cdot) \right)  \right)\\
       & = E^{\pi_M}_{x,\nu^*}\left( \gamma_{\l}\left(\o_{n-1}(\xi^*_{n-1}, \cdot) M \right) \cdot \beta_i^{y,n}(\chi_{\l})  \right)\\
       & = \beta_i^{y,n}(\chi_{\l}) \cdot \gamma_{\l}\left(E^{\pi_M}_{x,\nu^*}(\o_{n-1}(\xi^*_{n-1}, \cdot) M)\right)\\
       & = \beta_i^{y,n}(\chi_{\l}) \cdot \gamma_{\l}\left(E^{\pi_M}_{x,\nu^*}(\o_{n-1} M)\right) = \beta_i^{y,n}(\chi_{\l}) \cdot \gamma_{\l}(\pi_M),
   \end{align*}
   where the third equality follows from the fact that $\gamma_{\l}$ is affine, and the fourth one follows from the fact that given $\o_{n-1}$, $\o_{n-1}(\xi^*_{n-1}, \cdot)$ has mean $\o_{n-1}$. The proof of the claim is completed, as by arguments parallel to those given in the proof of Claim \ref{Claim Step 1} we have $P^{\pi_M}_{x,\nu^*} \left( \o_n(\xi_n^*, \cdot) = \chi\,|\, \{\o_n \in \R\} \right) = \beta_i^{y,n}(\chi_{\l}) \cdot \gamma_{\l}(\pi_M)$ as well. 
   \end{proof}
The above two adjusted proofs suffice, as the proofs of Claims \ref{Claim Step 3} and \ref{Claim Step 4} remain the same in the current setup. Thus, Lemma \ref{Lemma Distribution Equality} follows, as required. 

\subsubsection{Proof of Proposition \ref{Prop Mono 2-3}}

In the case where $x=0$, the recursive formula for $\Gamma_0^*$, with discount factor $\delta \in [0,1)$ takes the form:

\begin{align*}
     \Upsilon_{\delta} (\o) & = \sup_{(\beta_a, \rho_a) \in \mathcal{S}(\o)} \sum_{a \in A}\, \beta_a \cdot\bigg\lbrace (1-\delta) \cdot u + \delta  \cdot \Upsilon_{\delta} \circ M  \bigg\rbrace (\rho_a) \nonumber \\
    & = (\text{Cav}\, \phi_{\delta})(\o)
\end{align*}
where $\phi_{\delta}(\o) := (1-\delta) \cdot u(\o)+ \delta  \cdot (\Upsilon_{\delta} \circ M)(\o)$. Moreover, as Carath\'{e}odory's Theorem (see, e.g., Corollary 17.1.5 in \cite{Rock}) implies that for each $\o \in \Delta(K)$, the above supremum can be attained by a split to at most $k$ elements, we infer that $\Upsilon_{\delta}(\o)$ is the same for all sets $A$ with $|A|\geq k$. In particular, we may take $A=S$, where $S$ is the signal set in the Markovian persuasion game. However, for such a choice of $A$, the reduced MDP $\G_0^*$ agrees with the MDP reformulation of $MP(0)$ given in \ref{MDP of MP(x)}. Thus, we obtain that $ \Upsilon_{\delta} (\o) = v_{\delta} (\o)$ for every $\o \in \Delta(K)$, as required. 

\subsection{Proof of Theorem \ref{Thm RG-mono}}

By the same line of arguments as in Proposition \ref{Prop Mono 2-1} we obtain that $V^x_N (\pi_M)$ converges to $\g_1(\pi_M) \cdot V_{1-x}(\chi_1) + ... + \g_k(\pi_M) \cdot V_{1-x}(\chi_k)$ as $N \to \infty$, where $V^x_N(\cdot)$ is the value with respect to the $N$-Ces\`aro average payoff. Next, the fact that for every $N\geq 1$, the mapping  $x \mapsto V^x_N (\pi_M)$ is non-increasing in $x$ follows from the cheap-talk assumption, as at any stage, Player 1 can communicate a message which will split the beliefs of Player 2 to $\{\chi_1,...,\chi_k\}$. Thus, Player 1 can always make the coupling, which is the pillar of all monotonicity results in the paper. The fact that such a coupling works follows similar arguments to those given in the proof of Proposition \ref{Prop Mono 2-2}. As before, the combination of the convergence of $V^x_N (\pi_M)$ together with their monotonicity in $x$ completes the proof.


\begin{thebibliography}{9}

\bibitem{Galit}
Ashkenazi-Golan, G., Hernández, P., Neeman, Z., and Solan, E. (2023), Markovian persuasion with two states. \emph{Games Econom.\ Behav., }142, 292--314.

\bibitem{Aumann}
Aumann, R.J. and Maschler, M. (1995), \emph{Repeated Games with Incomplete Information}. With the collaboration of R.
Stearns,  MIT Press, \emph{Cambridge, MA}.

\bibitem{Bizzotto}
Bizzotto, J., R\"{u}diger, J., and Vigier, A. (2021). Dynamic persuasion with outside information.
American Economic Journal: Microeconomics, 13(1):179–194.

\bibitem{Blackwell}
Blackwell, D. (1951), Comparison of Experiments. In: \emph{Proceedings of the Second Berkeley Symposium on Mathematical Statistics and Probability, 1950}, pp. 93--102. University of California Press, Berkeley and Los Angeles, CA.

\bibitem{Ely}
Ely, J.C. (2017), Beeps. \emph{American Economic Review}, 107(1), 31--53.

\bibitem{Farhadi}
Farhadi, F. and Teneketzis, D. (2022), Dynamic Information Design: A Simple Problem on Optimal
Sequential Information Disclosure. \emph{Dyn.\ Games Appl.,} 12(2), 443--484.


\bibitem{Jackson Kalai}
Jackson, M.\ O., and Kalai, E. (1997), Social Learning in Recurring Games, \emph{Games Econ.\ Behav.,} 21(1-2), 102--134.


\bibitem{Kamenica}
Kamenica, E. and Gentzkow, M. (2011), Bayesian Persuasion. \emph{American Economic Review}, 101 (6), 2590--2615.


\bibitem{MP}
Lehrer, E. and Shaiderman, D. (2023), Markovian persuasion. \emph{Theoretical Economics}, to appear. \\
\href{https://econtheory.org/ojs/index.php/te/article/viewForthcomingFile/5372/42663/1}{https://econtheory.org/ojs/index.php/te/article/viewForthcomingFile/5372/42663/1}.

\bibitem{Lehrer}
Lehrer, E. and Sorin, S. (1992), A Uniform Tauberian Theorem in Dynamic Programming. \emph{Math.\ Oper.\ Res.\ }17, no. 2, 303--307.


\bibitem{Peres}
Levin, D. A. and Peres, Y. (2017), \emph{Markov Chains and Mixing Times}, second edition. With contributions by Elizabeth L. Wilmer. With a chapter on ``Coupling from the past'' by James G. Propp and David B. Wilson. American Mathematical Society, Providence, RI.

\bibitem{Lorecchio}
Lorecchio, C. (2022). Persuading crowds. UB Economics–Working Papers, 2022,
E22/434.

\bibitem{MSZ book}
Mertens, J.F., Sorin, S. and Zamir, S. (2015), \emph{Repeated games}. Cambridge: Cambridge University Press.

\bibitem{Renault}
Renault, J. (2006), The value of Markov chain games with lack of information on one-side. \emph{Math.\ Oper.\ Res.\ }31, no. 3, 433--648.

\bibitem{Solan}
Renault, J., Solan, E., and Vieille, N. (2017), Optimal Dynamic Information Provision. \emph{Games Econ.\ Behav.,} 104, 329--349.

\bibitem{SolanB}
Solan, E. (2022), \emph{A Course in Stochastic Game Theory.} Cambridge University Press. 

\bibitem{Rock}
Rockafellar, R. T. (1970), \emph{Convex Analysis}. Princeton University Press.

\end{thebibliography}
\end{document}